\documentclass[aps,pra,reprint,amsmath,amssymb,superscriptaddress]{revtex4-1}
\usepackage{amsmath}
\usepackage{times}

\usepackage[caption = false]{subfig}

\usepackage[colorlinks=true,linkcolor=blue,urlcolor=blue,citecolor=blue]{hyperref}
\usepackage{color}
\usepackage{amscd}
\usepackage{amsfonts}
\usepackage{amssymb}
\usepackage{graphicx, epstopdf}
\usepackage{tabularx}

\newlength\mytemplen
\newsavebox\mytempbox

\makeatletter
\newcommand\mybluebox{%
    \@ifnextchar[
       {\@mybluebox}%
       {\@mybluebox[0pt]}}

\def\@mybluebox[#1]{%
    \@ifnextchar[
       {\@@mybluebox[#1]}%
       {\@@mybluebox[#1][0pt]}}

\def\@@mybluebox[#1][#2]#3{
    \sbox\mytempbox{#3}%
    \mytemplen\ht\mytempbox
    \advance\mytemplen #1\relax
    \ht\mytempbox\mytemplen
    \mytemplen\dp\mytempbox
    \advance\mytemplen #2\relax
    \dp\mytempbox\mytemplen
    \colorbox{myblue}{\hspace{1em}\usebox{\mytempbox}\hspace{1em}}}

\begin{document}


\title{Quasiperiodic quantum heat engines with a mobility edge}

\author{Cecilia Chiaracane}
\affiliation{Department of Physics, Trinity College Dublin, Dublin 2, Ireland}
\author{Mark T. Mitchison}
\affiliation{Department of Physics, Trinity College Dublin, Dublin 2, Ireland}
\author{Archak Purkayastha}
\affiliation{Department of Physics, Trinity College Dublin, Dublin 2, Ireland}
\author{G\'eraldine Haack}
\affiliation{Universit\'e de Gen\`eve, Department of Applied Physics,Chemin de Pinchat 22, CH-1211 Geneve 4, Switzerland}
\author{John Goold}
\affiliation{Department of Physics, Trinity College Dublin, Dublin 2, Ireland}

\begin{abstract}
    
Steady-state thermoelectric machines convert heat into work by driving a thermally generated charge current against a voltage gradient. In this work, we propose a new class of steady-state heat engines operating in the quantum regime, where a quasiperiodic tight-binding model that features a mobility edge forms the working medium. In particular, we focus on a generalization of the paradigmatic Aubry-Andr\'e-Harper (AAH) model, known to display a single-particle mobility edge that separates the energy spectrum into regions of completely delocalized and localized eigenstates. Remarkably, these two regions can be exploited in the context of steady-state heat engines as they correspond to ballistic and insulating transport regimes. This model also presents the advantage that the position of the mobility edge can be controlled via a single parameter in the Hamiltonian. We exploit this highly tunable energy filter, along with the peculiar spectral structure of quasiperiodic systems, to demonstrate large thermoelectric effects, exceeding existing predictions by several orders of magnitude. This opens the route to a new class of highly efficient and versatile quasiperiodic steady-state heat engines, with a possible implementation using ultracold neutral atoms in bichromatic optical lattices. 

\end{abstract}
\date{\today}
\maketitle

\section{Introduction}
\label{sec:intro}

Scientific activity in the area of thermal machines has been boosted in recent years by the increasing importance that society is placing on sustainable energy. The inevitable tendency towards miniaturization and the importance of recycling waste heat provide strong motivations to consider microscopic thermal machines, for which quantum effects can become relevant~\cite{kosloff2014, mark} and may even be exploited (see, for example, Refs.~\cite{Scully2011,Brunner2014,Correa2014,Killoran2015,Biele2017,Ghosh2017,YungerHalpern2019,Klatzow2019}). Thermoelectric engines, in particular, do not rely on macroscopic moving parts. Instead, they convert heat into power through nonequilibrium steady-state currents of microscopic particles, e.g., electrons or atoms, flowing between two reservoirs. Unfortunately, bulk thermoelectrics are generally quite inefficient~\cite{benenti2017}. This drawback, together with the unprecedented level of control achieved in nanotechnology, has fuelled both experimental and theoretical research to identify and characterize new nanoscale systems to be harnessed as efficient thermal engines~\cite{josefsson2018, giazotto2014, Zebarjadi_2012, Bell:2008AA, dresselhaus2007new, Benjamin2017}.

A central concept in thermoelectric energy conversion is energy filtering: in order to obtain a strong thermoelectric response, it is necessary to allow only particles in a finite energy window to flow \cite{MahanS1996, whitney2014most}. This effect is generally realized either by engineering the thermodynamic variables of the reservoirs~\cite{Bosisio2014,benenti2017} or by tuning the transport characteristics of the sample so that it displays an energy-dependent transmission probability. In this work, we follow the latter approach by exploiting the spectral characteristics of the central system; in particular, we use a mobility edge as an energy filter. A mobility edge separates localized eigenstates from extended ones in systems with an energy-dependent localization transition. The most famous example is the three-dimensional Anderson model, where random disorder drives localization~\cite{anderson1958,Semeghini2015}, leading to a diverging thermoelectric response in the vicinity of the mobility edge~\cite{sivan1986, yamamoto2017}. However, this enhancement does not appear in lower spatial dimensionalities, where all states are localized by infinitesimal disorder, independently of their energy~\cite{gangof4}. 

Yet if the random disorder is replaced by a quasiperiodic potential, incommensurate with the underlying periodicity of the lattice, a localization transition with a mobility edge can occur even in one dimension. A paradigmatic example is the Aubry-Andr\'e-Harper (AAH) model~\cite{aubry1980analyticity, harper1955single}. The AAH model shows a phase transition from a completely delocalized phase to a completely localized phase as the strength of the quasiperiodic potential is increased~\cite{aubry1980analyticity}. At the critical point, both the spectrum of the AAH model and the eigenfunctions have a fractal nature~\cite{Pandit83,Kohmoto1987}, a property that is also of interest to mathematicians~\cite{Last1994, hiramoto}. The standard AAH model features no mobility edge in any of the phases. However, adding perturbations to the AAH model, e.g., by allowing beyond-nearest-neighbour hopping or, as in the present work, by deforming the on-site potential, leads in many cases to the occurrence of a mobility edge~\cite{Rossignolo2019, dasSarma_bichromatic, GAAH1, biddle2011localization, biddle2010localization, biddle2009localization, boers2007mobility, dasSarma_early2,dasSarma_early1, settino2017}. 

Quasiperiodic systems, with and without mobility edges, and with tunable interaction strength, have been realized in experiments on ultracold atoms trapped by two optical lattices with different wavelengths~\cite{Roati2008,schreiber2015, Bloch2017, Bloch2018, Bloch2019}. These systems formed the basis of recent investigations into many-body localization and the effect of interactions in the presence of a mobility edge~\cite{Bloch2019,GAAH4, GAAH5, Garg2017, machine_learning_mobility_edge}. The peculiar transport properties of non-interacting quasiperiodic systems have recently been characterized~\cite{purkayastha2019,purkayastha2018,GAAH6,vkv2017}, and their possible applications as rectifiers have been highlighted~\cite{rectification2,rectification3}. On the other hand, a completely different set of experiments have made tremendous progress in realizing two-terminal transport measurements with ultracold atoms~\cite{Esslinger2018,Esslinger2017,Esslinger2016,Esslinger2014,Esslinger2013}. 

Motivated by these works, here we propose another interesting application of quasiperiodic systems: namely, as the working medium of a quantum thermal machine. To that end, we perform the first characterization of the thermoelectric properties of a one-dimensional quasiperiodic system. We focus on the generalized AAH (GAAH) model recently introduced in Ref.~\cite{GAAH1}, for which an exact analytical expression for the mobility edge is known. The spectral properties~\cite{GAAH2,GAAH3} and the open-system particle transport properties of the GAAH model~\cite{GAAH6} have been previously investigated.  Focusing on the experimentally relevant linear-response regime, we compute the transport coefficients using a Landauer-B\"uttiker approach and use these to quantify the performance of a GAAH heat engine as a function of temperature. We show how the remarkable spectral properties of quasiperiodic systems give rise to a versatile and efficient quantum thermal machine. In particular, the position of the mobility edge can be tuned by modifying a single parameter in the Hamiltonian, while the presence of ballistic states above the mobility edge leads to significant power output even in the limit of large system size. We also demonstrate that the combination of ballistic transport and a mobility edge enhances efficiency when compared to a homogeneous wire. Finally, we show that the physics described here is not only limited to our chosen model, but is also expected to hold true in more general cases.

The rest of this article is organized as follows. In Section~\ref{sec:thermal_machines}, we describe the general setup and recall the Landauer-B\"uttiker formalism for characterizing transport in the linear response regime. The GAAH model and its spectral characteristics are discussed in Section~\ref{sec:GAAH_model}. Our main results are presented and discussed in Section~\ref{sec:results}, where we compute the transport properties and thermodynamic performance of the GAAH machine for specific examples. We summarize and conclude in Section~\ref{sec:conclusions}. 

\section{Autonomous thermoelectric machines}
\label{sec:thermal_machines}

\subsection{General setup}

\begin{figure}
\centering
\includegraphics[scale=0.56]{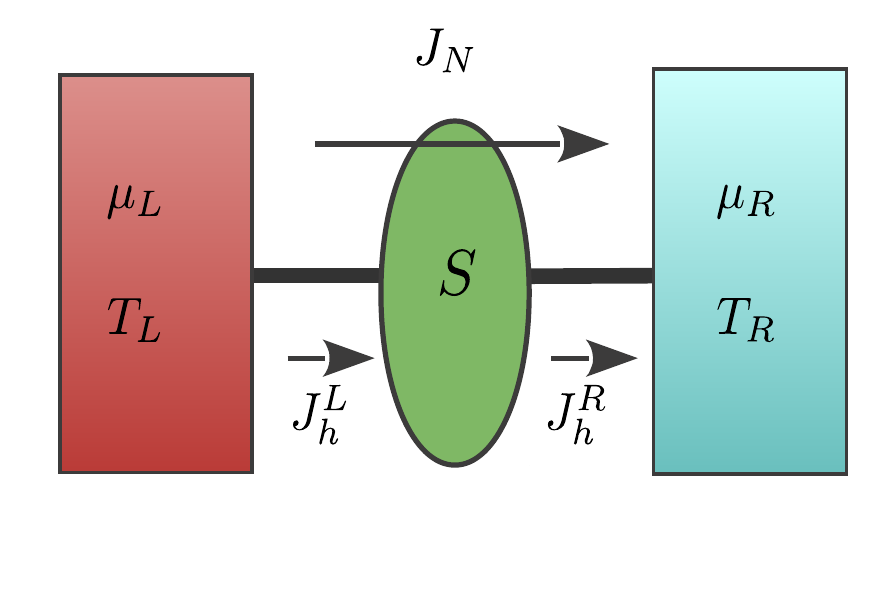} 
\caption{Schematic of the thermoelectric heat engine. We engineer the thermodynamic properties of the reservoirs in such a way to use the thermal current to drive electrons against a chemical potential bias.}
\label{fig:draw1}
\end{figure}

In this work, we exclusively focus on two-terminal devices that function as autonomous thermal machines. A thermoelectric engine may best be understood by example. Consider a situation typical in mesoscopic physics, where two metallic leads are connected to a central region, as illustrated in Fig.~\ref{fig:draw1}. The two leads are electron reservoirs assumed to be in thermal equilibrium, with well defined chemical potentials $\mu_L$, $\mu_R$ and temperatures $T_L$, $T_R$. In the long-time limit, the central system reaches a nonequilibrium steady state characterized by nonzero charge and heat currents. These currents are not generated by any external drive, but entirely by the temperature bias or the difference in the chemical potentials, so the machine is said to be autonomous.  

We assume without loss of generality that the left reservoir is hotter than the right one, i.e., $T_L>T_R$. The temperature difference induces an electrical current $J_e = e J_N$ from the left reservoir into the right one, with $e$ the electron charge and $J_N$ the particle current. The associated energy current is denoted by $J_E$. As a consequence of these currents, in each unit of time a quantity of heat $J^h_L$ flows out of the left reservoir while $J^h_R$ flows into the right reservoir. From the first law of thermodynamics and the conservation of charge and energy, the expressions for the heat currents are found to be~\cite{callen,yamamoto2015}
\begin{equation}
    J_h^{\nu} = J_E - \mu_{\nu} J_N,
\end{equation}
for $\nu = L,R$. The difference between the heat currents corresponds to the power developed by moving electrons from a low chemical potential to a higher one, viz. 
\begin{equation}
    P =  J_h^L - J^R_h = J_N \Delta \mu = J_e \Delta V ,
\end{equation}
where $\Delta \mu = \mu_R-\mu_L$ is the chemical potential difference and $\Delta V = \Delta \mu/e$ is the applied voltage. The system behaves as a heat engine whenever $P>0$, in which case the efficiency is given by
\begin{equation}
    \eta = \dfrac{P}{J_h^L} = 1 - \dfrac{J_h^R}{J_h^L}.
\end{equation}
The expression is the same as for a standard cyclic thermal engine, and it is bounded from above by the corresponding Carnot efficiency $\eta_C = 1 - T_R/T_L$.  

\subsection{Thermodynamics in linear response}
\label{sec:linear_response}

From here on, we consider the linear-response regime, where the differences between chemical potentials $\Delta \mu=\mu_L-\mu_R$ and temperatures $\Delta T=T_L-T_R$ are small compared to their averages. This regime is relevant for numerous experimental platforms, ranging from semiconductor~\cite{Snyder2008, shakouri2011} and molecular~\cite{diventra2011} electronics to ultracold atoms~\cite{Esslinger2013}. The extension of the present work beyond linear response is straightforward but significantly more involved and will form the topic of a future publication.

In the linear-response regime, the currents can be expressed as linear combinations of the generalized forces or affinities driving transport~\cite{callen,Houten1992,de2013non,benenti2017}. This relation is compactly represented via the Onsager matrix~\cite{Onsager1931} as
\begin{align}
\left( \begin{array}{c} J_e \\ J_q \end{array} \right) & = \mathcal{L}\left( \begin{array}{c} \Delta\mu/eT \\ \Delta T/T^2 \end{array} \right) , \quad \mathcal{L}
=\begin{pmatrix} L_{11} & L_{12} \\ L_{21} & L_{22} \end{pmatrix},
\label{eq:onsanger}
\end{align}
with $L_{12} = L_{21}$ if the system satisfies time-reversal symmetry. 
The electrical conductance $G$, the thermal conductance $K$ and the Seebeck factor (or thermopower) $S$ are defined as
\begin{align}
	G&= \left. \left( \dfrac{J_{e}}{\Delta V} \right)\right\vert_{\Delta T=0}=\dfrac{L_{11}}{T} , \label{eq:Gcoeff}\\ 
	K&=\left. \left( \dfrac{J_{h}}{\Delta T} \right) \right\vert_{J_e=0} =\dfrac{1}{T^2}\frac{\det\mathcal{L}}{L_{11}} , \label{eq:Kcoeff} \\	
		S&=-\left.\left( \frac{\Delta V}{\Delta T} \right)\right \vert_{J_e=0} = \frac{1}{T}\frac{L_{12}}{L_{11}}. \label{eq:Scoeff}	
\end{align}
These three transport coefficients fully characterize heat-to-work conversion in the non-equilibrium steady state. 

For a time-reversal symmetric system in the linear-response regime, the maximum thermodynamic efficiency $\eta_{max}$ reachable by the device can be written in terms of a single dimensionless parameter $ZT= \dfrac{GS^2T}{K}$ as~\cite{goldsmid2010intro}
\begin{equation}
\eta_{max} = \eta_C \frac{\sqrt{ZT +1} - 1}{\sqrt{ZT + 1} + 1}.
\end{equation}
Larger values of $ZT$ correspond to higher efficiencies. In particular, as $ZT \rightarrow \infty$, the Carnot efficiency is obtained, $\eta_{max} \rightarrow \eta_C$, which usually implies zero power output. In order to evaluate the capability of a system under practical conditions, we thus focus on another quantity: the efficiency at maximum power~\cite{vanDenBroeck2005}, which can be expressed as
\begin{equation}
\eta(P_{max})  = \dfrac{\eta_C}{2} \dfrac{ZT}{ZT + 2}.
\label{eq:etaPmax}
\end{equation}    
This tends to $\eta_C/2$ for $ZT \rightarrow \infty$, which is equivalent to the Curzon-Ahlborn (CA) bound~\cite{Curzon1975} in the linear-response regime.

The figure of merit $ZT$ is an important index to categorize thermoelectrics~\cite{Snyder2008,Bell:2008AA} (even though it may over- or underestimate the performance of the engine outside of the linear-response regime). Most current thermoelectric devices work with $ZT \approx 1$ and it is often stated that $ZT\approx 3$ would be required in order to compete with with alternative technologies~\cite{benenti2017}. Suggestions to increase $ZT$ include the use of nanostructures~\cite{dresselhaus2007new,Zebarjadi_2012,bian2007cross} and the breaking of time-reversal symmetry~\cite{benenti2011thermodynamic, sanchez2015, sanchez2015B, haack2019}. For quantum-coherent transport, however, perhaps the simplest way to increase performance is to use an energy filter, as discussed in the following section. 

\subsection{Landauer-B{\"u}ttiker formalism}
\label{sec:LB} 

We now specifically focus on the coherent regime, where the particles, here spinless electrons, crossing the central region undergo elastic scattering events without dissipation of energy. The currents in this case are given by the Landauer-B{\"u}ttiker formalism via the integrals 
 \begin{align}
 \label{LB_charge_current}
 	J_{e}&= \frac{2e}{h}\int dE\tau(E)[f_L(E)-f_R(E)],  \\
 	 \label{LB_energy_current}
 	J^{\nu}_{h}&=\frac{2}{h}\int dE(E-\mu_{\nu})\tau(E)[f_L(E)-f_R(E)],
 \end{align}
where the factor 2 is due to the spin degeneracy, and $f_{\nu}(E)=\{1 + \exp[(E - \mu_{\nu})/k_B T_{\nu}]\}^{-1}$ is the Fermi-Dirac distribution of bath $\nu=L,R$, with $h$ and $k_B$ the Planck and Boltzmann constants, respectively. The transmission function $\tau(E)$ encodes the probability for an electron at energy $E$ to tunnel from the left to the right reservoir through the central region. 

In the linear-response regime, we may Taylor-expand the Fermi-Dirac distributions around reference thermodynamic variables $\mu = \mu_R$ and $T = T_R$. Comparing the result with Eq.~\eqref{eq:onsanger} gives
$L_{11}=e^2TI_{0}$, $L_{12}=L_{21}=eTI_{1}$ and $L_{22}=TI_{2}$, where
\begin{align}
\label{buttik}
	 I_{k}&=\frac{2}{h}\int dE(E-\mu)^{k}\tau(E)[-f'(E)], \\
	 f'(E) &=\frac{\partial f}{\partial E}=-[4k_{B}T\cosh^{2}(E-\mu)/2k_{B}T]^{-1}  \nonumber.      
\end{align}
Note that $f'(E)$ is an even function centered around $\mu$ with a width of order $k_{B}T$. 

\begin{figure}
\centering
\includegraphics[scale=0.57]{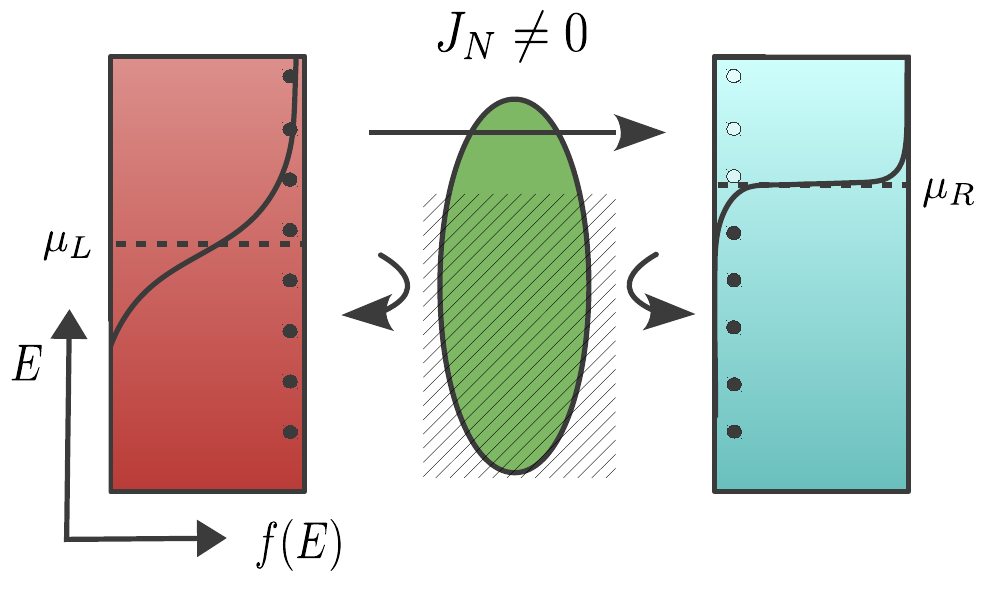} 
\caption{An efficient thermoelectric device can be obtained through the use of an energy filter in the central system, blocking the transport at certain energies. The temperature bias drives particle (hole) transport above (below) the chemical potential, leading  to zero net electric current in the presence of particle-hole symmetry. Finite electric current and, consequently, output power are instead obtained by differentiating the dynamics of the particles at energies above and below the chemical potential.}
\label{fig:draw2}
\end{figure}

Let us now examine the effect of an energy filter. The central system behaves as an energy filter whenever the transmission function $\tau(E)$ is zero for energies above or below a certain value. This is a mechanism to break the particle-hole symmetry that would otherwise impede thermoelectric power generation. Indeed, in the presence of particle-hole symmetry, heat is transported both by particles above the chemical potential and by holes below the chemical potential. The corresponding charge currents of the particles and the holes compensate each other, leading to zero net power output. As demonstrated in Fig.~\ref{fig:draw2}, blocking transport in the working medium within a certain energy range allows charge to flow only in one direction, i.e., against the voltage gradient. Mathematically, the effect of an energy filter is seen by using Eq.~\eqref{buttik} to write the thermopower as 
\begin{equation}
    S=\frac{1}{eT}\frac{\int^{\infty}_{-\infty} dE(E-\mu)\tau(E)[-f'(E)]}{\int^{\infty}_{-\infty} dE\tau(E)[-f'(E)]}.
\end{equation}
Given that $f'(E)$ is an even function of the energy, it is clear that the Seebeck factor will vanish whenever the transmission probability is also an even function. Breaking electron-hole symmetry in the transmission probability is therefore crucial to achieve a finite thermoelectric response. 

\subsection{Transmission function for one-dimensional wires}
\label{sec:transmission} 

As is evident from the above discussion, the fundamental object governing the behavior of an autonomous engine is the transmission function $\tau(E)$, which depends on the microscopic details of the central system and its coupling to the reservoirs. In this work, we focus on the situation where a one-dimensional (1D) tight-binding model of non-interacting fermions is connected at the boundaries to two non-interacting reservoirs. This is described by a generic Hamiltonian  $\hat{H}=\hat{H}_{S}+\hat{H}_{SE}+\hat{H}_{E}$. The Hamiltonian of the system is given by
\begin{equation}
\label{wireHam}
\hat{H}_{S}=\sum_{i=1}^{N-1}  t (\hat{a}^{\dagger}_{i}\hat{a}_{i+1}+{\rm h.c})+ \sum_{i=1}^{N} V_i\hat{a}^{\dagger}_{i}\hat{a_{i}},
\end{equation}
where $V_i$  is the on-site energy of site $i$,  $t$ is the tunnelling constant and  $\hat{a}_i$ is the fermionic annihilation operator of site $i$. The reservoirs are described by quadratic fermionic Hamiltonians with infinitely many degrees of freedom. The combined Hamiltonian of both baths is given by 
\begin{align}
\hat{H}_{E}=\sum_{\nu=L,R}\sum_{k}E_{k\nu}\hat{d}_{k\nu}^{\dagger}\hat{d}_{k\nu},
\end{align}
 where $E_{k\nu}$ are the single-particle eigenenergies of leads and $\hat{d}_{k\nu}$ are annihilation operators for the corresponding eigenmodes. We assume a bilinear system-reservoir coupling of the form 
\begin{align}
\hat{H}_{SE}=\sum_{k}(t_{kL}\hat{a}^{\dagger}_{1}\hat{d}_{kL}+t_{kR}\hat{a}^{\dagger}_{N}\hat{d}_{kR}+ {\rm h.c})
\end{align} 
where $t_{kL}$ and $t_{kR}$ describe the amplitude for electrons to tunnel from the left and right leads onto the wire. Note that first site of the system is coupled to the left lead (denoted by the subscript $L$) and the last site of the system is coupled to the right lead (denoted by the subscript $L$).  Each bath is described by a spectral function 
\begin{align}
\mathfrak{J}_{L/R}(E)=2\pi\sum_{k}|t_{k L/R}|^2\delta(E-E_{k L/R}).
\end{align}
We make the wide-band limit (WBL) approximation, taking spectral functions that are identical and independent of energy: $\mathfrak{J}_{L}(E)=\mathfrak{J}_{R}(E)=\gamma$. 

For this situation, the transmission function $\tau(E)$ can be exactly calculated using non-equilibrium Green functions (NEGF)~\cite{ryndyk2016, datta1997}. To this end, we first  write the system Hamiltonian as
\begin{align}
\hat{H}_S = \sum_{ij}H_{ij}\hat{a}_i^\dagger\hat{a}_j,
\end{align}
where $H$ is a symmetric tridiagonal matrix with diagonal entries $\{V_i\}$ and off-diagonal entries equal to $t$. The retarded single particle NEGF of the set-up is given by the $N \times N$ matrix $ G(E)=M^{-1}(E)$, with 
\begin{align}
M(E)=[E\mathbb{I}- H - \Sigma_{L}(E) - \Sigma_R(E)],
\end{align}
where $\mathbb{I}$ is the $N$-dimensional identity matrix and $\Sigma_{L,R}(E)$ are $N$-dimensional matrices representing the self-energies of the baths. The transmission function is then given by $\tau(E)= \text{Tr}\{\Gamma_{L}(E) G^{\dagger}(E) \Gamma_{R}(E) G(E)\}$~\cite{Meir1992}, where the level-width functions are defined as $\Gamma_{L,R}(E)=i(\Sigma_{L,R}^\dagger(E)- \Sigma_{L,R}(E))$. For our set-up, the matrices $\Sigma_{L}(E)$ and $\Sigma_{R}(E)$ have only one non-zero element each, given by $\left[\Sigma_{L}(E)\right]_{11}=\left[\Sigma_{R}(E)\right]_{NN} = - i\gamma/2$. The expression for the transmission function thus simplifies to
\begin{align}
\tau(E) = \gamma^2 \vert G_{1N}(E)\vert^2= \frac{\gamma^2}{\vert\textrm{det}\left[M(E)\right]\vert^2}.
\end{align}

\section{The generalized Aubry-Andr\'e-Harper model}
\label{sec:GAAH_model}

A spectacular realization of the energy filtering mechanism discussed in Sec.~\ref{sec:LB} is the mobility edge associated with the metal-insulator transition of the Anderson model~\cite{anderson1958,sivan1986, yamamoto2017}.  Here, random disorder localizes only the low-energy part of the spectrum, while high-energy states remain extended. This leads to an asymmetric transmission function and hence a diverging thermopower in the vicinity of the mobility edge, which separates the localized, insulating states from the extended, conducting ones. The Anderson metal-insulator transition occurs in three spatial dimensions \cite{anderson1958}, while in lower dimensions, and in the absence of interparticle interactions, all states are localized in the thermodynamic limit~\cite{gangof4}. 

Here, instead, we focus on the thermoelectric properties of quasiperiodic systems, which have a disordered, yet \textit{non-random}, potential that leads to localization. Remarkably, quasiperiodic systems can exhibit a mobility edge even in one spatial dimension, unlike the case of random disorder.

The particular quasiperiodic model that we choose is the generalized Aubry-Andr\'e-Harper (GAAH) model given by following on-site potential~\cite{GAAH1}
\begin{equation}
V^{GAAH}_i = \dfrac{2 \lambda \cos(2 \pi b i + \varphi)}{1 - \alpha \cos(2 \pi b i + \varphi)}.
\label{eq:gaapot}
\end{equation}
Here, $\lambda$ indicates the strength of the potential, $\varphi$ is a phase that shifts the origin of the potential, $b$ is an irrational number and $\alpha\in (-1,1)$. Choosing $b$ to be an irrational number makes the cosine incommensurate with the underlying periodicity of the lattice. Of course, in experiments, truly irrational numbers do not exist. Nevertheless, the model is always realized for a finite system of size $N$ if $b = p/q$ is taken as a rational number, with $q>N$ and $p,q$ coprime, such that the potential has a different value on every site. 

For $\alpha=0$, the GAAH model reduces to the Aubry-Andr\'e-Harper (AAH) model~\cite{aubry1980analyticity}. In the AAH model,  the quasiperiodic nature of the potential gives the spectrum a fractal structure and leads to a delocalization-localization transition depending just on $\lambda$~\cite{aubry1980analyticity}. The critical point of the transition in the AAH model occurs at $\lambda=t$. For $\lambda<t$ all single-particle eigenstates are completely delocalized, leading to ballistic transport, while for $\lambda>t$, all single-particle eigenstates are localized. For $\lambda=t$, the states are multifractal, and lead to counter-intuitive anomalous transport behavior as recently demonstrated~\cite{purkayastha2019,purkayastha2018,vkv2017}. The AAH model does not feature a mobility edge in any phase.

\begin{figure}
\centering
\includegraphics[scale=0.33]{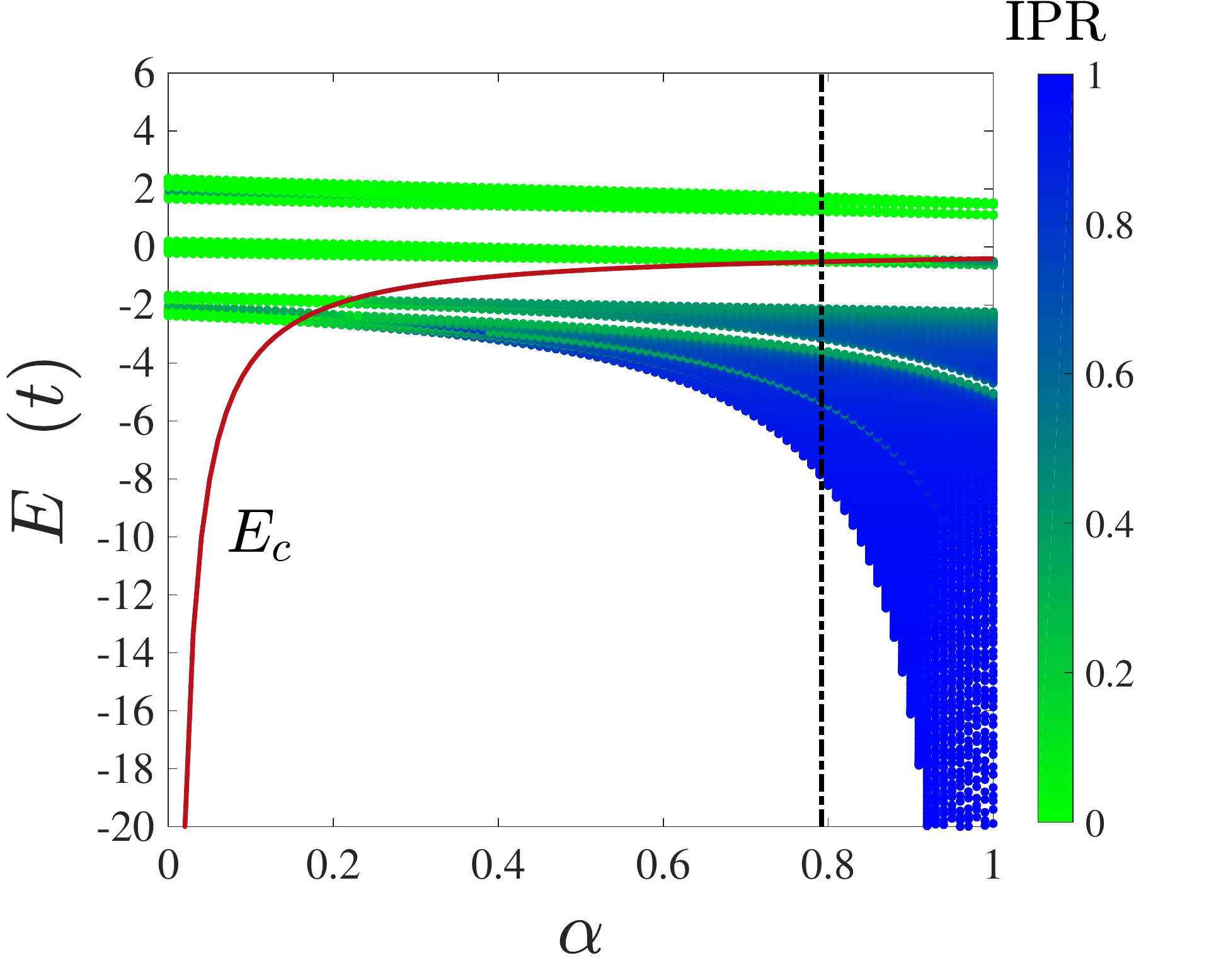} 
\caption{Eigenenergy spectra of GAAH systems with $b = (\sqrt{5} + 1)/2$, $\lambda  = -0.8$ and $\varphi = 0$ as a function of $\alpha$, for a chain of $N = 987$ sites. The IPR of the corresponding eigenstate is shown by a color map, with green for extended, blue for completely localized states. The red line represents the mobility edge $E_c$ given by Eq.~\eqref{eq:me}, which separates localized from delocalized states. The dashed black line indicates the value of $\alpha$ chosen in Sec.~\ref{sec:results}.}
\label{fig:spectrum}
\end{figure}

For $\alpha\neq 0$, the GAAH model features a mobility edge in energy, separating the regions of completely delocalized and localized states in the same spectrum. Most interestingly, for this model, the energy of the mobility edge has been shown analytically to be~\cite{GAAH1}
\begin{equation}
E_c = \dfrac{1}{\alpha}\text{sign}(\lambda)(\vert t \vert - \vert \lambda \vert).
\label{eq:me} 
\end{equation}
Thus, the position of the mobility edge can be tuned by changing $\lambda$ and $\alpha$, leading to a much richer phase diagram. The high temperature nonequilibrium phase diagram of the GAAH model has been explored in Ref.~\cite{GAAH6}. The precise knowledge of the position of the mobility edge for given values of $\lambda$ and $\alpha$ makes the GAAH model ideal for investigation of low temperature thermoelectric properties in 1D quasiperiodic systems. 

One important difference between the mobility edges appearing in the GAAH model and in the three-dimensional Anderson model is in the nature of the conducting states. The conducting states in the case of the GAAH model support ballistic transport~\cite{GAAH6}, whereas those in the three-dimensional Anderson model support diffusive transport. As we will see, this has a major effect on the power output of our quasiperiodic quantum thermal machine.

In order to quantify the localization properties of the spectrum as a function of energy, we use the inverse participation ratio (IPR)
\begin{equation}\label{IPR}
\text{IPR}(E_{n})=\sum_{\ell}|\Phi_{\ell n}|^{4},
 \end{equation}
where $\Phi_{\ell n}$ is the single-particle eigenfunction with eigenenergy $E_{n}$, evaluated at lattice site $\ell$. For localized states, the IPR is close to its maximum value of unity and does not scale with system size, while for extended states it is of order $N^{-1}$, i.e. vanishingly small in the thermodynamic limit $N\to\infty$. For multifractal states, IPR$\sim N^{-p}$, with $0<p<1$. Fig.~\ref{fig:spectrum} shows the energy spectrum and corresponding IPR of the GAAH model for various values of $\alpha$ at a chosen value of $\lambda$. For our numerical calculations, here and henceforth we choose $b=(\sqrt{5} + 1)/2$ to be the golden mean~\cite{GAAH1}. 

 

\section{Results}
\label{sec:results}

\subsection{Transmission function}

\begin{figure}
\centering
\subfloat[]{\includegraphics[scale=0.24]{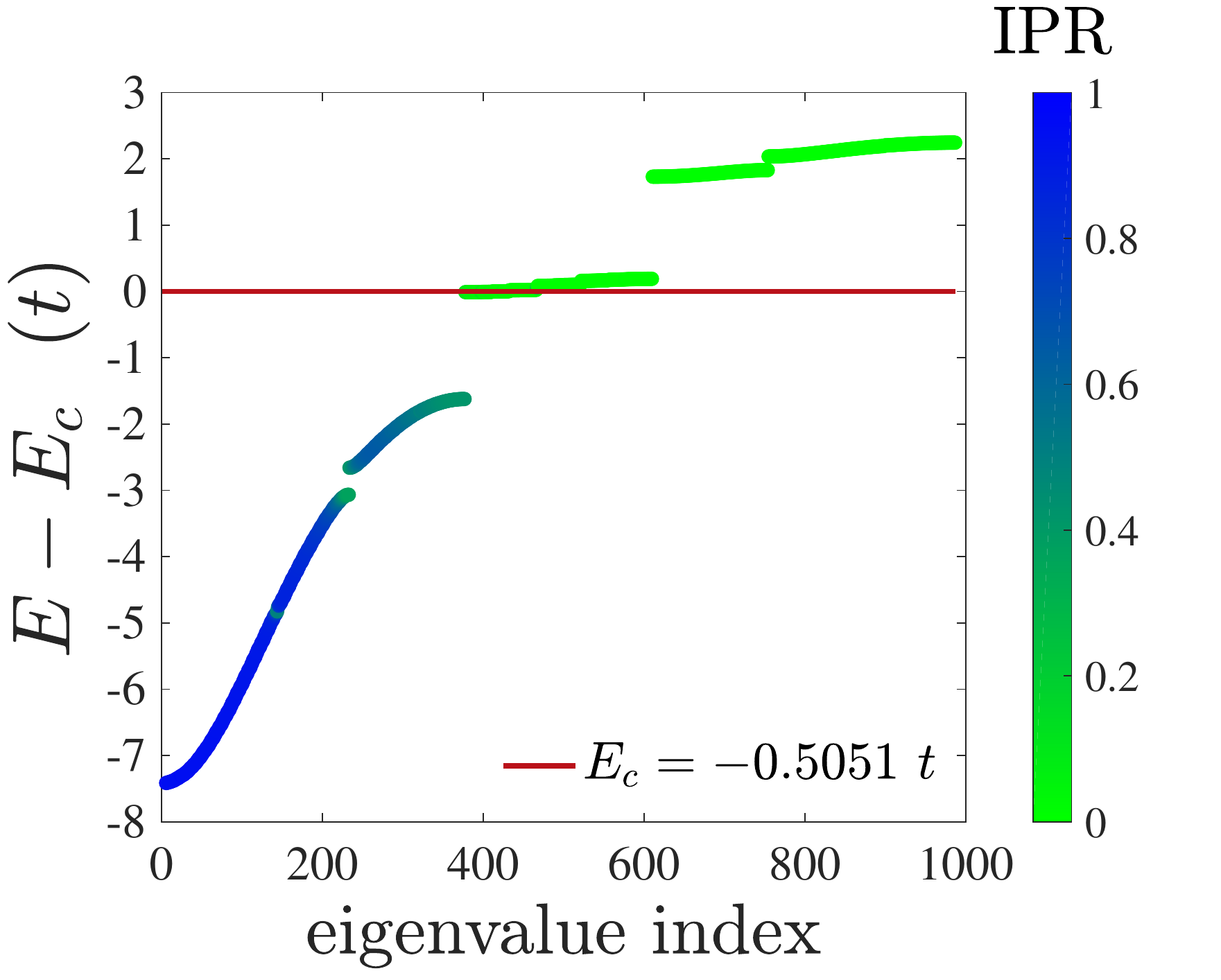} \label{fig:config1}}
\subfloat[]{\includegraphics[scale=0.24]{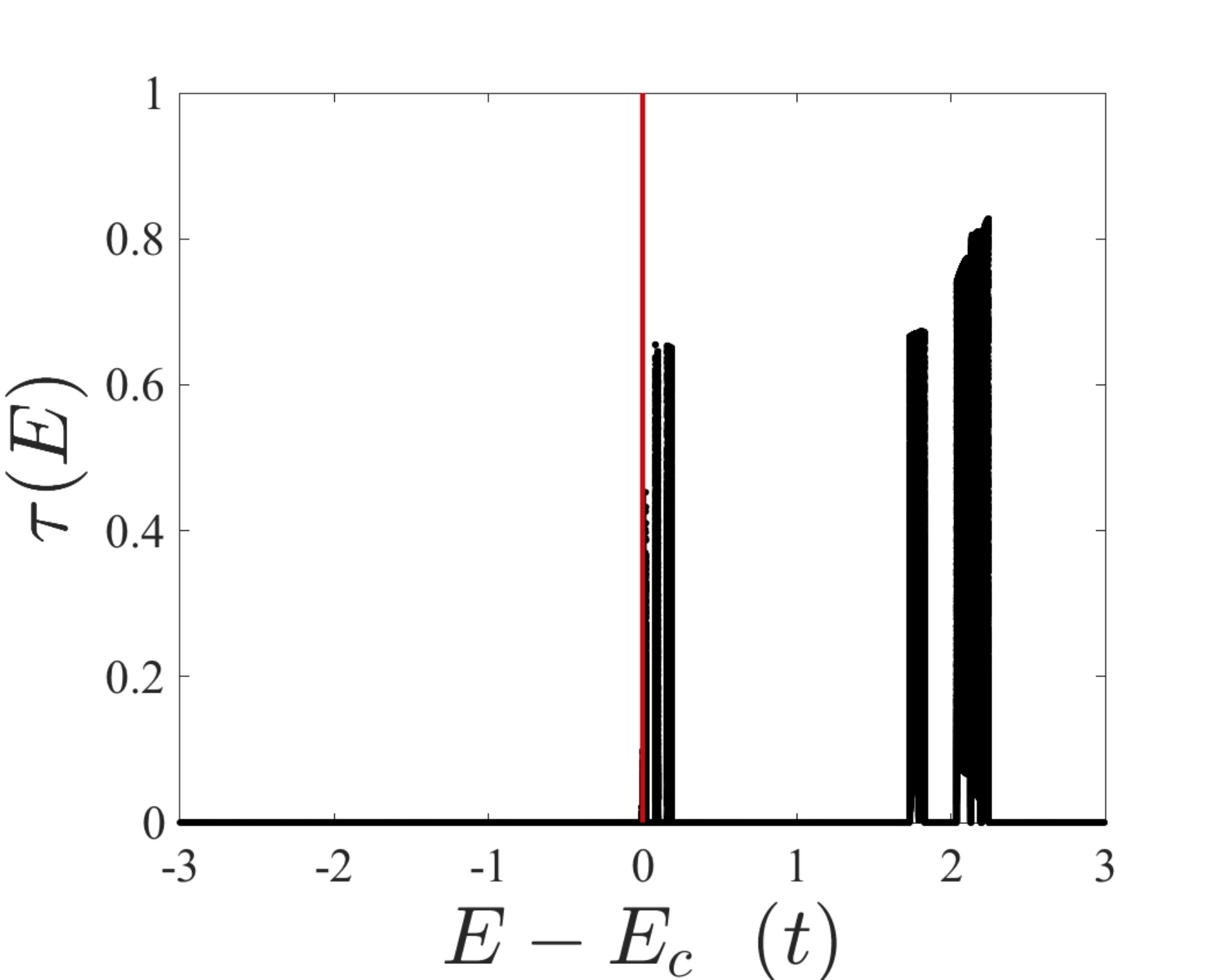} \label{fig:tau1} }\\
\subfloat[]{\includegraphics[scale=0.24]{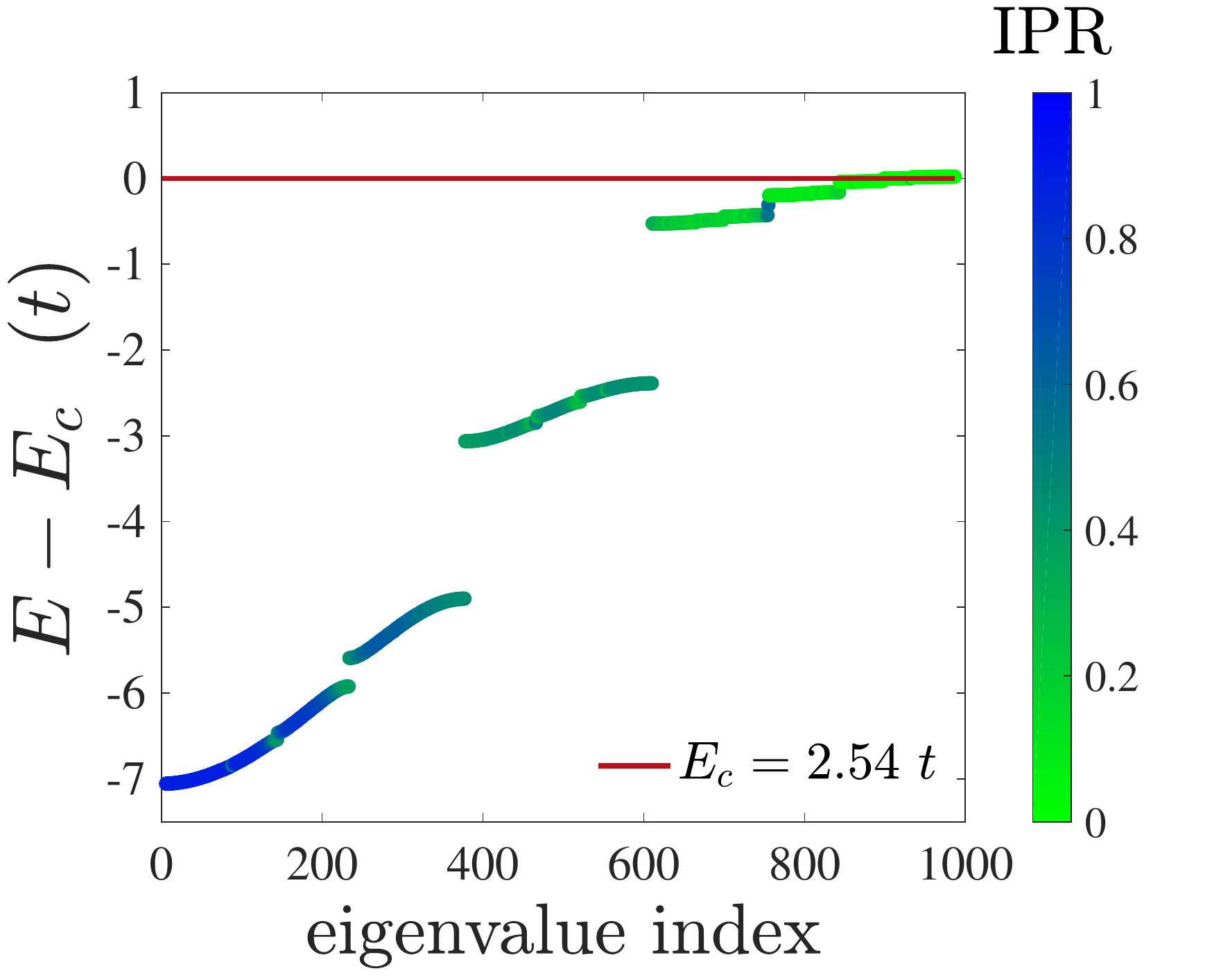} \label{fig:config2}}
\subfloat[]{\includegraphics[scale=0.24]{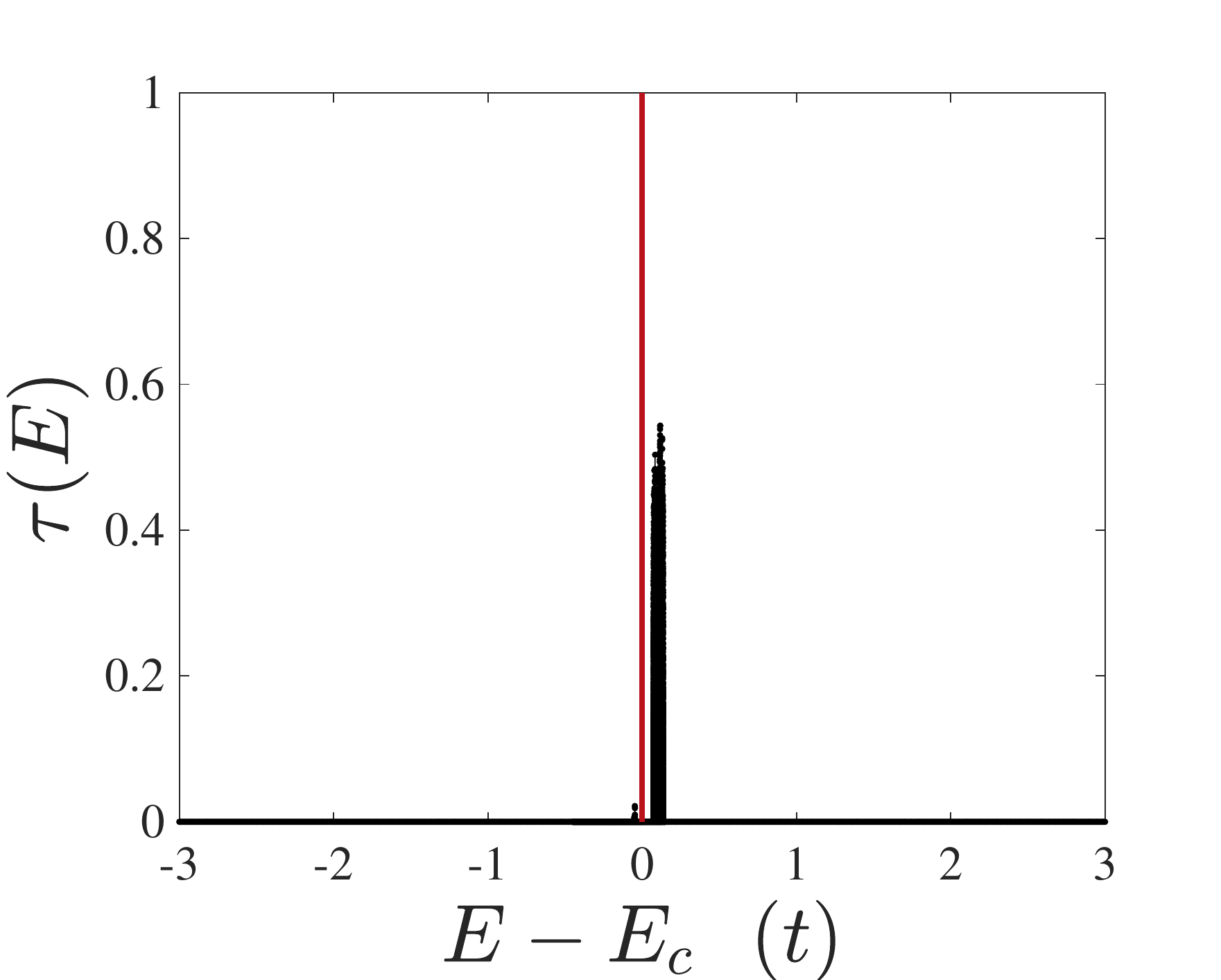} \label{fig:tau2}} \\
\caption{Spectra for single GAAH wires of length $N = 987$, generated with (a) $\lambda = -0.8 \ t, \ \alpha = 0.792, \ \varphi = 0$ (dashed vertical line in Fig.~\ref{fig:spectrum}), and (c) $\lambda = -1.4 \ t, \ \alpha = 0.330, \ \varphi = 0$. The mobility edge is shown by the red line. (b)-(d) The transmission functions associated respectively to the first and second configuration, averaged over 40 values of the phase  $\varphi$. Conduction is clearly possible only at energies that support extended eigenstates. \label{fig:sys}}
\end{figure}

The pivotal calculation for our results is the transmission function, which is independent of the temperature and the chemical potential of the reservoirs. In Fig.~\ref{fig:sys} we display the spectrum of the system and the corresponding transmission function for two different pairs of values of $\lambda$ and $\alpha$. We see that the mobility edge and the clusters of ballistic states lying above it generally give rise to a highly asymmetric transmission profile, which is conducive to a large thermoelectric response.  The choice of the two parameters in the model, moreover, gives control over the structure of the spectrum, determining the position of the mobility edge and the number of ballistic states above it.

A general property of quasiperiodic 1D systems is that the spectrum has fractal properties \cite{hiramoto}, which are reflected in the fine-grained structure of the transmission function \cite{GAAH6}. While these fractal properties depend on the exact choice of the quasiperiodic potential, the asymmetry of the transmission function and the occurrence of bands of ballistic states are generic to many 1D quasi-periodic systems with a mobility edge. As we show in Sec.~\ref{sec:taubox}, it is this generic behavior that governs thermoelectric properties, irrespective of further details.

At this point, a note on terminology is in order. For simplicity, we refer to the clusters of ballistic states lying above the mobility edges as ``bands'' in the following. Strictly speaking, these groups of states do not satisfy the usual definition of a band, because they do not tend to a continuum in the thermodynamic limit in the rigorous mathematical sense, due to their fractal structure. Nevertheless, as discussed above and shown in Sec.~\ref{sec:taubox}, this structure has little effect on thermodynamic properties such as efficiency, thus we make no strict distinction in terminology.

Since the GAAH model has a large parameter space, we focus on a single, representative example rather than performing an exhaustive study. In what follows, we consider the particular configuration displayed in Fig.~\ref{fig:config1}, corresponding to the dashed vertical slice in Fig.~\ref{fig:spectrum} with $\alpha = 0.792$, for a chain of 987 sites. The mobility edge $E_c$ sits within a group of closely packed eigenvalues, with several other ballistic bands. We work in a regime of intermediate system-bath coupling, $\gamma = t$. As shown in  Appendix~\ref{app:system_bath_coupling}, modifying $\gamma$ merely rescales the currents without qualitatively affecting the transport behaviour. In the following, we use this transmission function to analyze the thermoelectric properties of the GAAH wire in different temperature regimes, via the transport coefficients given by Eqs.~\eqref{eq:Gcoeff}--\eqref{eq:Scoeff}. All quantities shown in this section are obtained numerically and averaged over the phase $\varphi$ in Eq.~\eqref{eq:gaapot}.

\subsection{Low-temperature transport properties}

\begin{figure}
\centering
\subfloat[]{\includegraphics[scale=0.24]{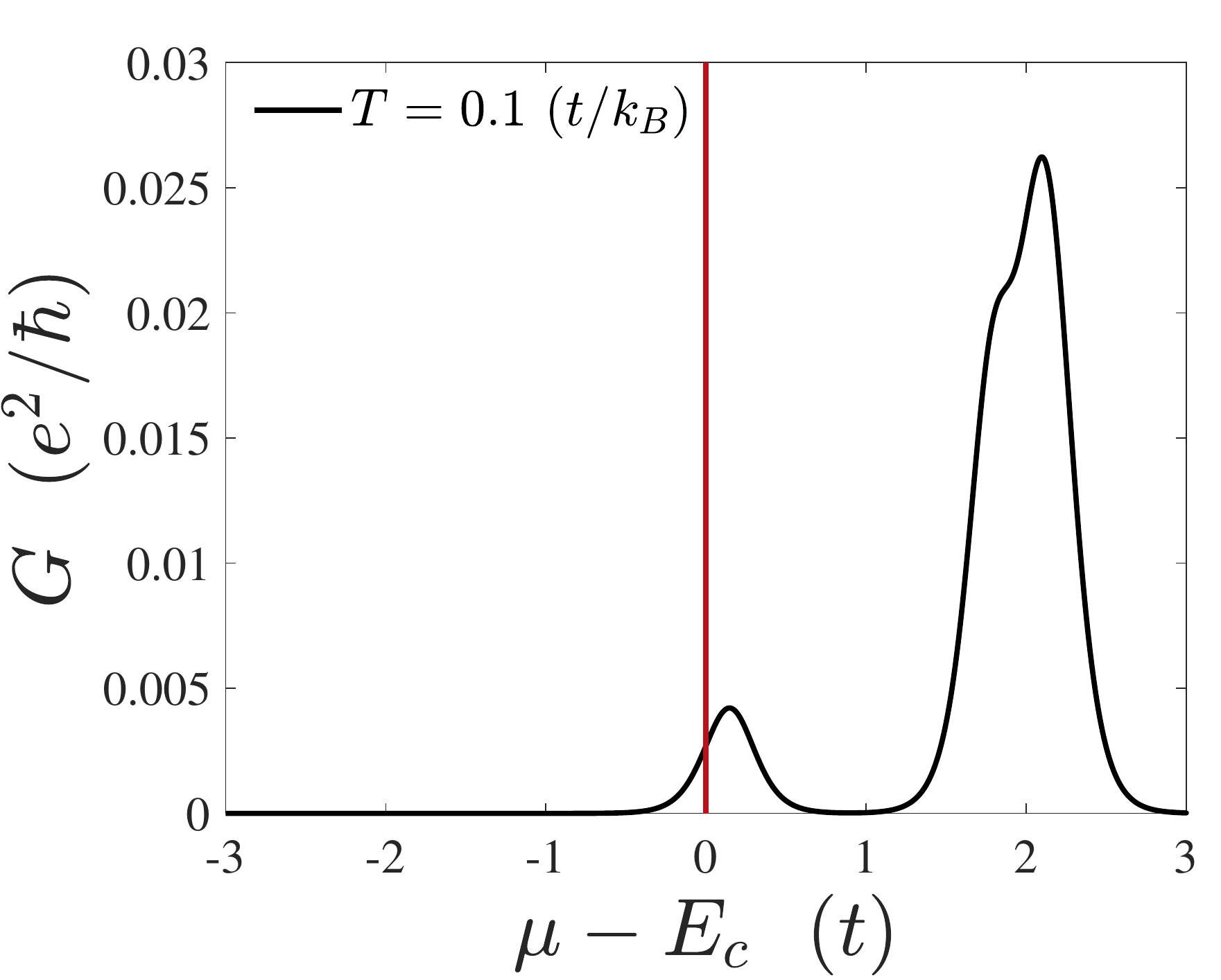} \label{fig:G}}
\subfloat[]{\includegraphics[scale=0.24]{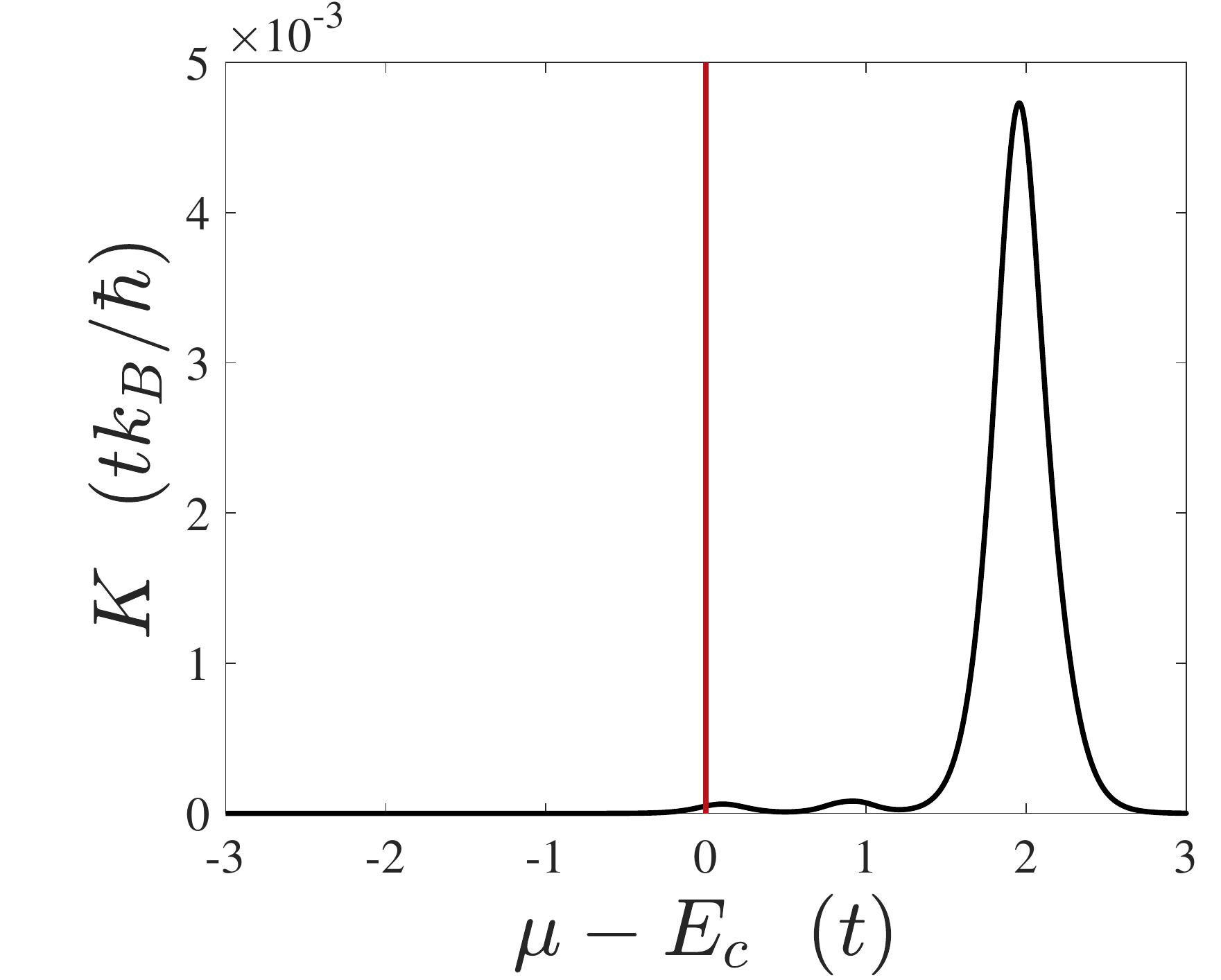} \label{fig:K}}
\caption{(a) Electric conductance and (b) thermal conductance as a function of chemical potential at fixed temperature $T=0.1(t/k_B)$.\label{fig:G_and_K}
}
\label{fig:tcoeff}
\end{figure}

We begin by studying the low-temperature behavior, choosing $T = 0.1 \ (t/k_B)$. This temperature regime is relevant for experiments involving ultracold atoms in optical lattices~\cite{Esslinger2018} and allows to clearly distinguish the non-trivial spectral structures reflected in the behaviour of the transport coefficients.

\begin{figure}[b]
\centering
\subfloat[]{\includegraphics[scale=0.24]{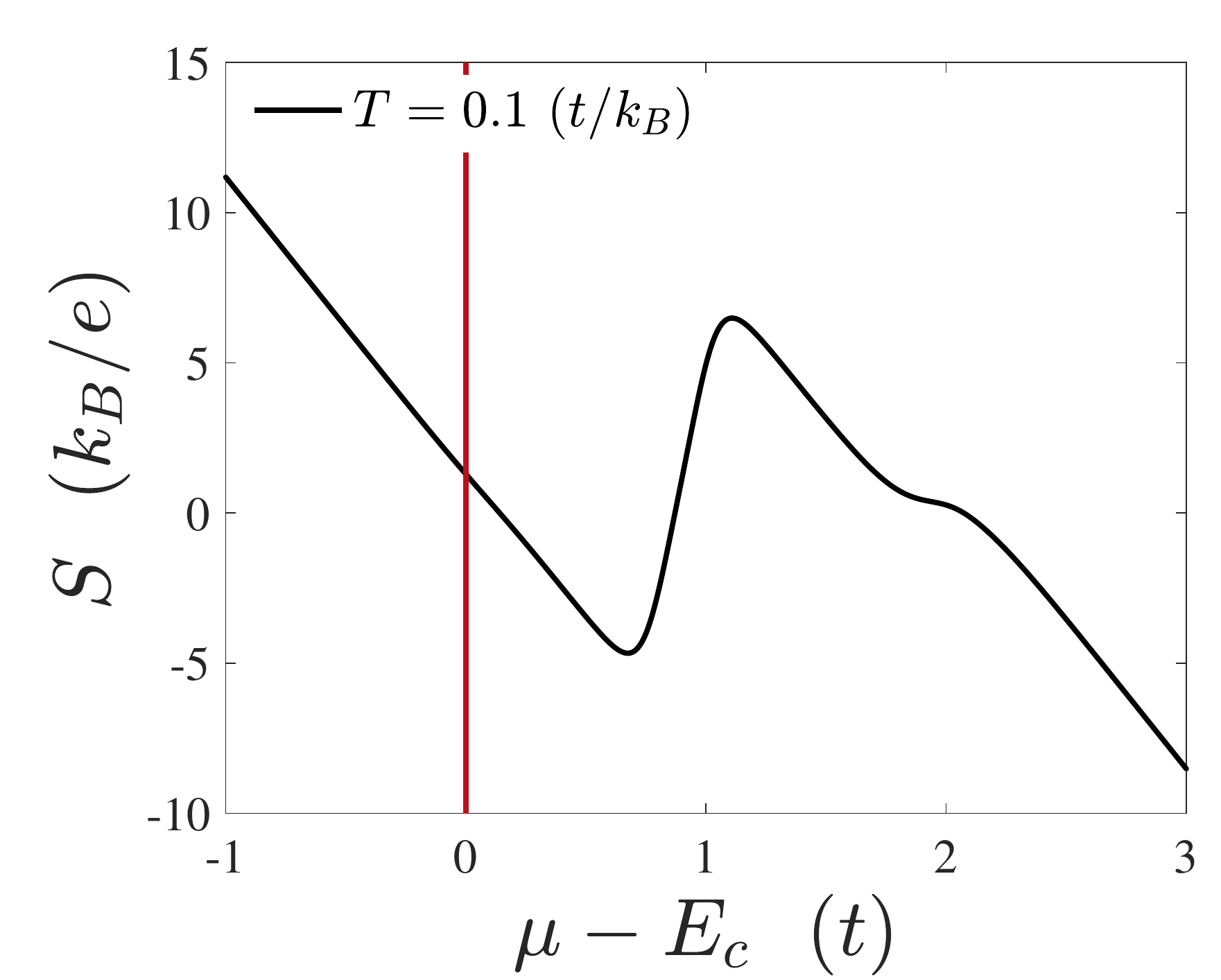} \label{fig:S}}
\subfloat[]{\includegraphics[scale=0.24]{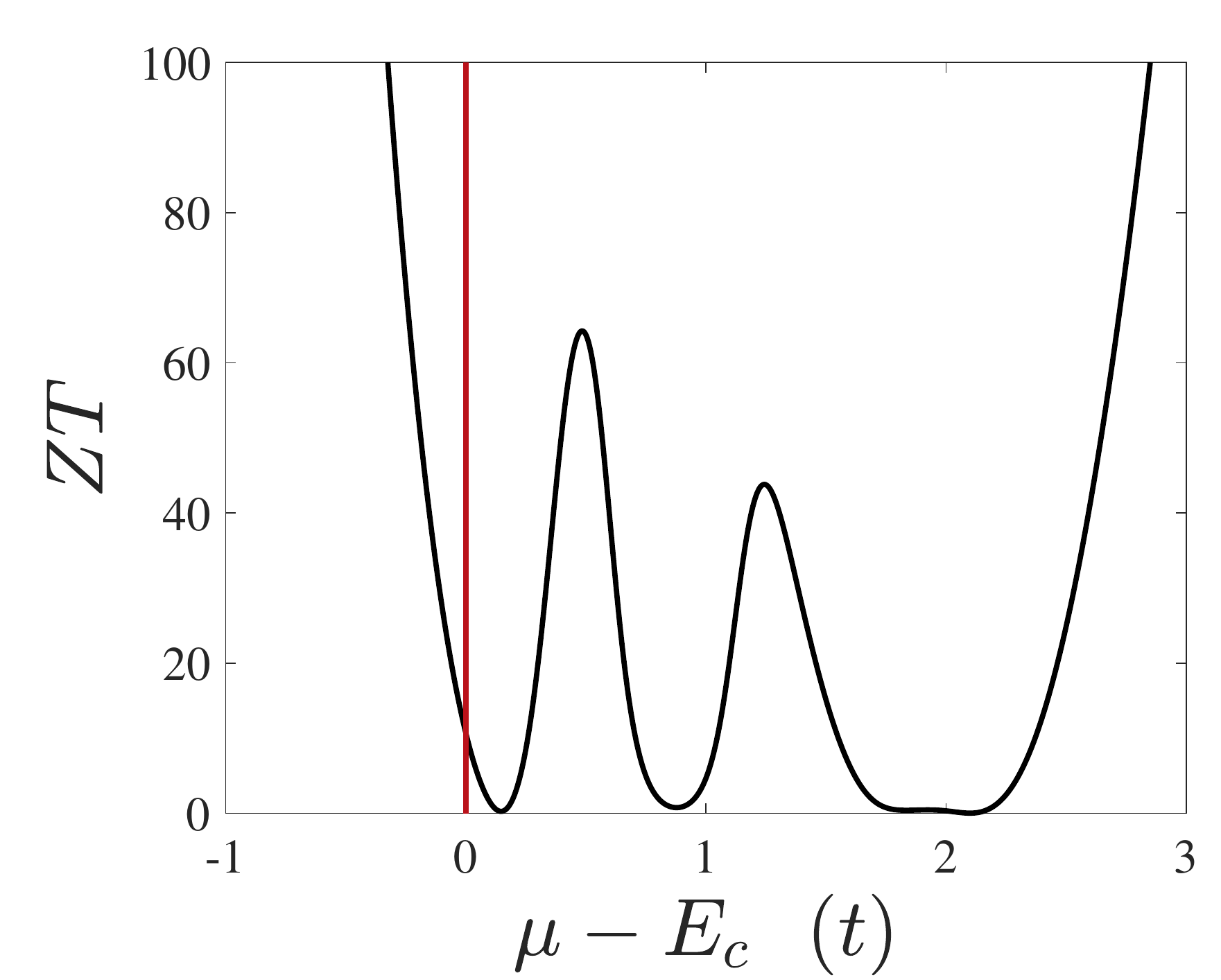} \label{fig:ZT} }
\caption{(a) Seebeck factor and (b) thermoelectric figure of merit as a function of chemical potential at fixed temperature $T=0.1(t/k_B)$.}
\end{figure}

We observe in Fig.~\ref{fig:G_and_K} that the electrical and thermal conductances closely follow the structure of the transmission function, with significant transport occurring only within the conducting bands around and above the mobility edge. The most dramatic effect due to the energy filter is evident in the Seebeck coefficient $S$ plotted in Fig.~\ref{fig:S}, which assumes finite values around the mobility edge. We also notice the magnitude of $S$ rising when the chemical potential is tuned far below or above the mobility edge. Even when $\mu$ lies on the insulating side, some of the delocalized states participate in transport because of the non-zero temperature, generating a small but finite conductance. As the mobility of the electrons decreases, the voltage necessary to stop their flux increases, leading to a large Seebeck factor according to Eq.~\eqref{eq:Scoeff}. In the region far above $E_c$, the charge carriers flow in the opposite direction to the heat carriers, leading to negative values for $S$.

We note that this behaviour of the transport coefficients can be reproduced by an approximate analytical calculation --- valid in the weak-coupling limit --- based on the localization properties of the energy eigenstates, as detailed in Appendix~\ref{app:analytical}. This analysis implies that similar thermoelectric characteristics will occur in other quasiperiodic systems displaying a mobility edge. 

The figure of merit $ZT$ also exhibits a divergence below the mobility edge, as shown in Fig.~\ref{fig:ZT}. This yields an extremely efficient thermal machine, yet in a region of negligible electrical conductance and thus vanishing power. Features more interesting for the realization of a useful device are instead visible when the chemical potential is tuned above the mobility edge. In this region, the engine has finite conductance, while the asymmetry of the transmission function gives rise to a figure of merit $ZT\approx 10$ just above the mobility edge. We observe, moreover, two higher peaks of $ZT \approx 60$ and $ZT\approx  40$ corresponding respectively to the upper and lower edges of the first and second ballistic bands above the mobility edge. Such values of $ZT$ correspond to efficiencies far exceeding those recorded in recent experiments~\cite{josefsson2018}.

\subsection{Efficiency at maximum power}

From the study of low-temperature transport, it is clear that by tuning the chemical potential it is possible to obtain an extremely efficient autonomous thermal machine at finite power output. In order to study the machine's performance more systematically, we now focus on the conditions for generating the maximum power. In the linear-response regime with fixed $\Delta T$, the power is maximized when $\Delta \mu = eS \Delta T/ 2$~\cite{benenti2017}. Since this value depends on the chemical potential through $S$, our goal is to find the best thermal machine, or, equivalently, the optimal $\mu$ in order to obtain the maximum power output.

\begin{figure}
\centering
\subfloat[]{\includegraphics[scale=0.24]{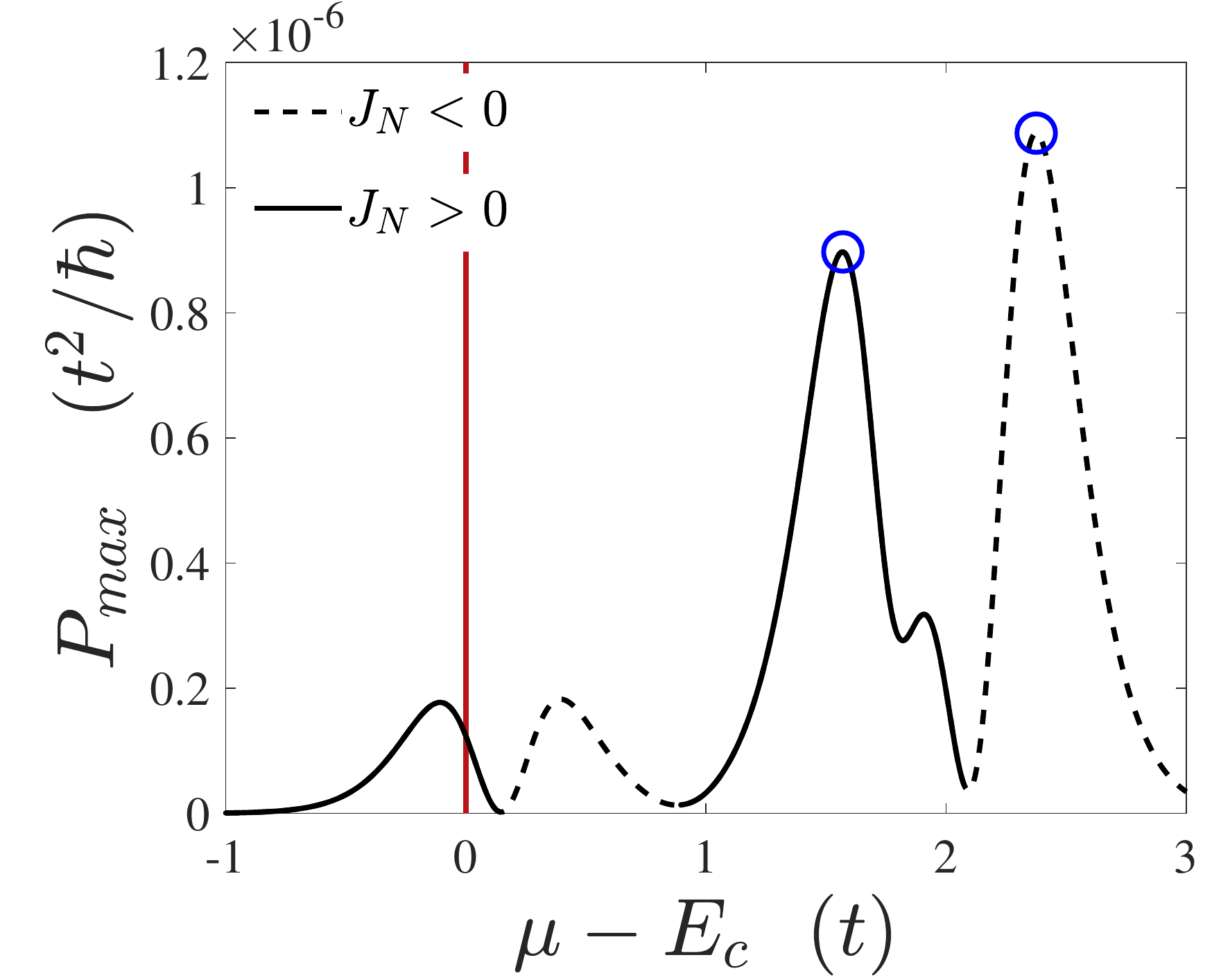} \label{fig:Pmax} }
\subfloat[]{\includegraphics[scale=0.24]{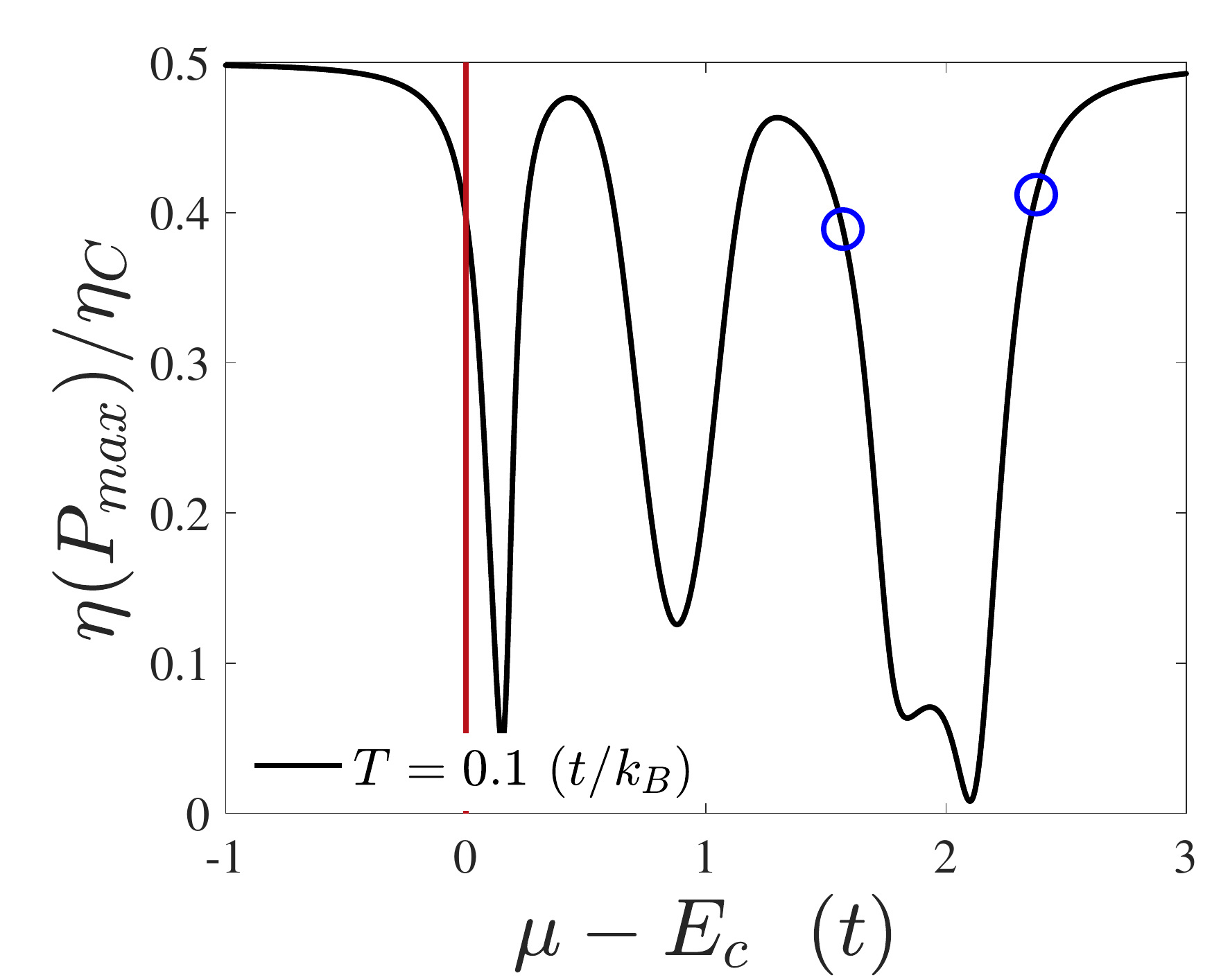} \label{fig:etaofPmax}}
\caption{(a) Maximum power and (b) efficiency at maximum power as a function of chemical potential at fixed temperature $T = 0.1$ and bias $\Delta T = 0.01(t/k_B)$. Blue circles mark the points of absolute maximum power.}
\label{fig:Pmaxeta}
\end{figure}

In Fig.~\ref{fig:Pmaxeta}, we plot the maximum power and corresponding efficiency \eqref{eq:etaPmax} as a function of $\mu$, at the fixed temperature $T = 0.1$ and bias $\Delta T = 0.01(t/k_B)$. We distinguish two cases according to whether the charge current is positive (left to right) or negative (right to left) according to our conventions. In the former case, the temperature gradient drives particle transport above the chemical potential, leading to power extraction for $\mu_R >\mu_L$. In the latter case, the thermal gradient causes holes below the chemical potential to migrate from left to right, which generates power so long as $\mu_L>\mu_R$. 

Two points that are particularly suitable for the realization of the thermal machine are marked with blue circles in Fig.~\ref{fig:Pmaxeta}: one in the region of positive $J_N$, the other for negative $J_N$. Here, the machine produces the highest values of electric power, with an efficiency reaching $\eta \approx  0.4 \eta_C$.  The strong thermoelectric response of the system at these two points is due to the lowermost and uppermost edges of the ballistic bands, respectively. Indeed, at low temperatures, it seems preferable to exploit the band edges rather than the mobility edge, since the power is significantly lower in the vicinity of $E_c$.

\subsection{Effect of increasing temperature}

\begin{figure}
\centering
\subfloat[]{\includegraphics[scale=0.24]{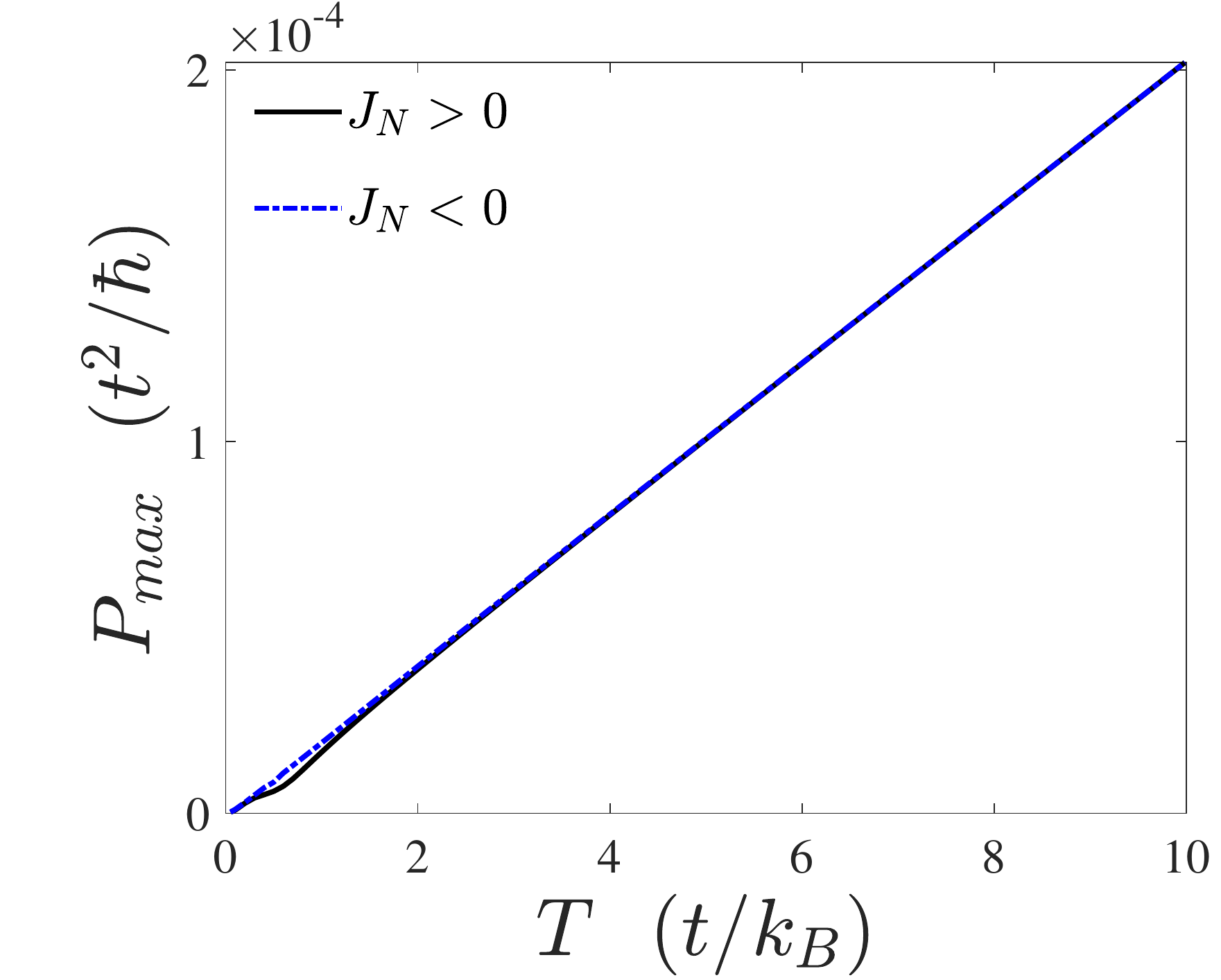} \label{fig:P1} }
\subfloat[]{\includegraphics[scale=0.24]{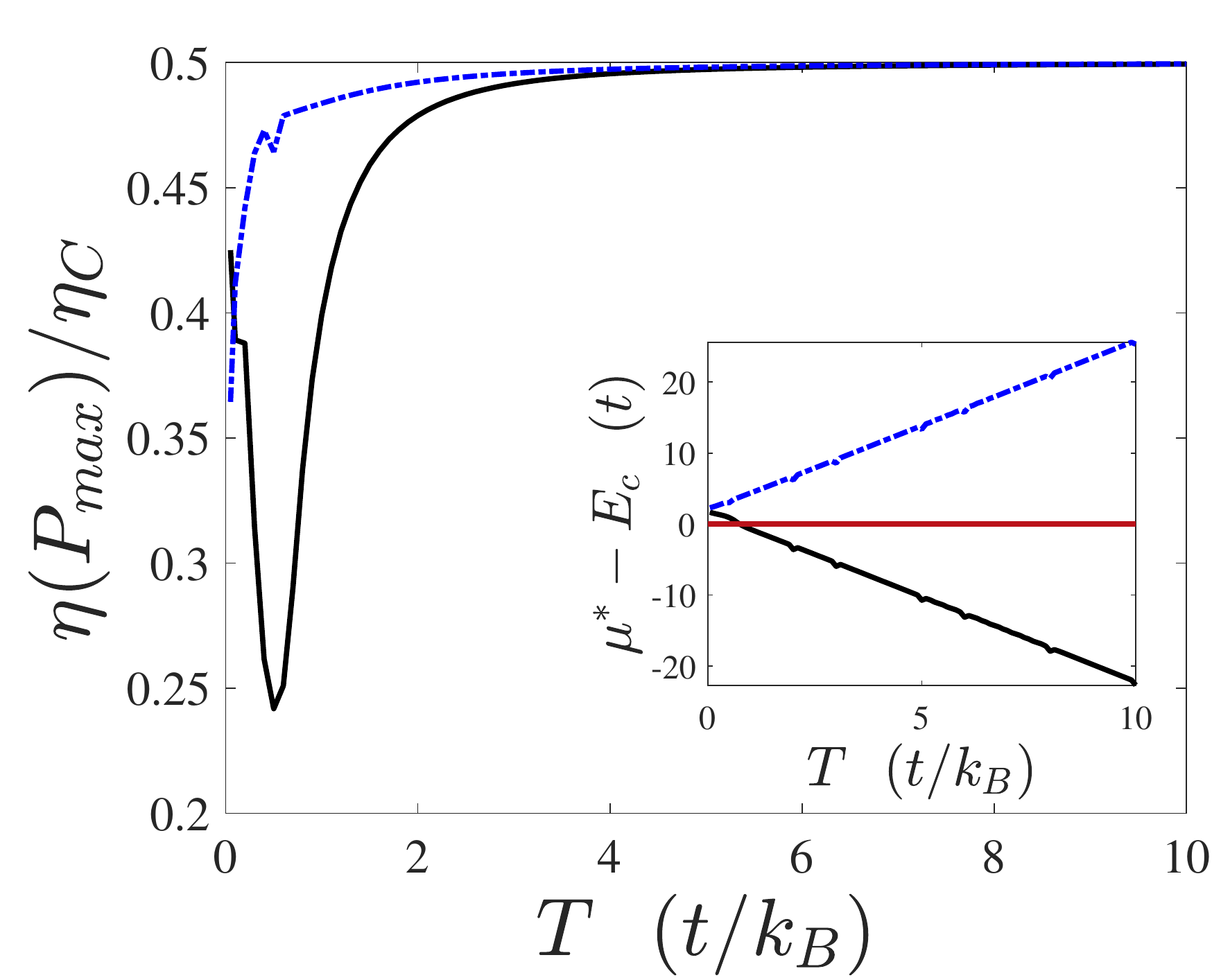} \label{fig:eta1}} 
\caption{(a) Absolute maximum power and (b) corresponding efficiency at maximum power as a function of temperature with $\Delta T/T = 0.1$. The solid black lines show results obtained by optimizing the power only over values of $\mu$ that give rise to a particle current flowing from the hot to the cold bath, $J_N > 0$. The dashed blue lines are instead obtained by restricting the maximization to $J_N<0$. The chemical potential yielding this maximum power is shown in the inset.}
\label{fig:perf}
\end{figure}

In this section, we explore the performance of the quasiperiodic machine at higher temperatures. In the linear-response regime, the temperature fixes the width of $f'(E)$, which determines the energy window centred on $\mu$ within which transport takes place. As $T$ increases, the gaps between the bands are no longer resolved and the sharp features of $G$ and $K$ displayed in Fig.~\ref{fig:G_and_K} are broadened and reduced in magnitude. As a result of this thermal broadening, the conductance is non-vanishing even for $\mu< E_c$ and the thermopower exhibits a weaker slope.

In order to meaningfully compare the thermoelectric performance of the GAAH wire at different temperatures, we vary $T$ while fixing the ratio $\Delta T/T = 0.1$, thus also ensuring that we remain in the linear-response regime. For each value of $T$ and $\Delta T$, we find the chemical potential, $\mu^*$, and bias, $\Delta \mu = eS(\mu^*) \Delta T/ 2$, that maximize power output. As before, we distinguish situations where $J_N>0$ and $J_N<0$ --- corresponding to heat transport by particles or holes, respectively --- performing a separate maximization for each case.

\begin{figure}[b]
\centering
\subfloat[]{\includegraphics[scale=0.24]{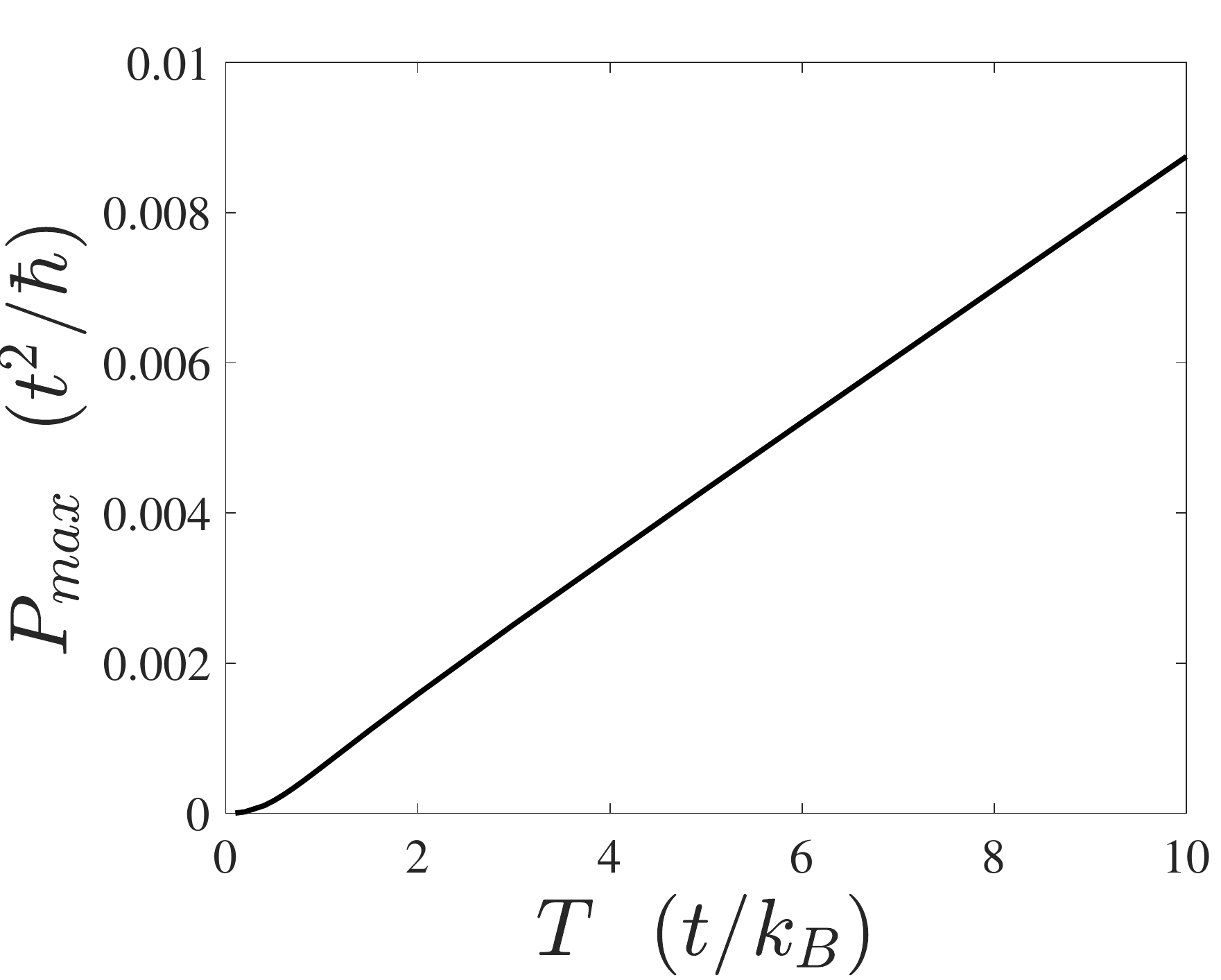} \label{fig:P_tb} }
\subfloat[]{\includegraphics[scale=0.24]{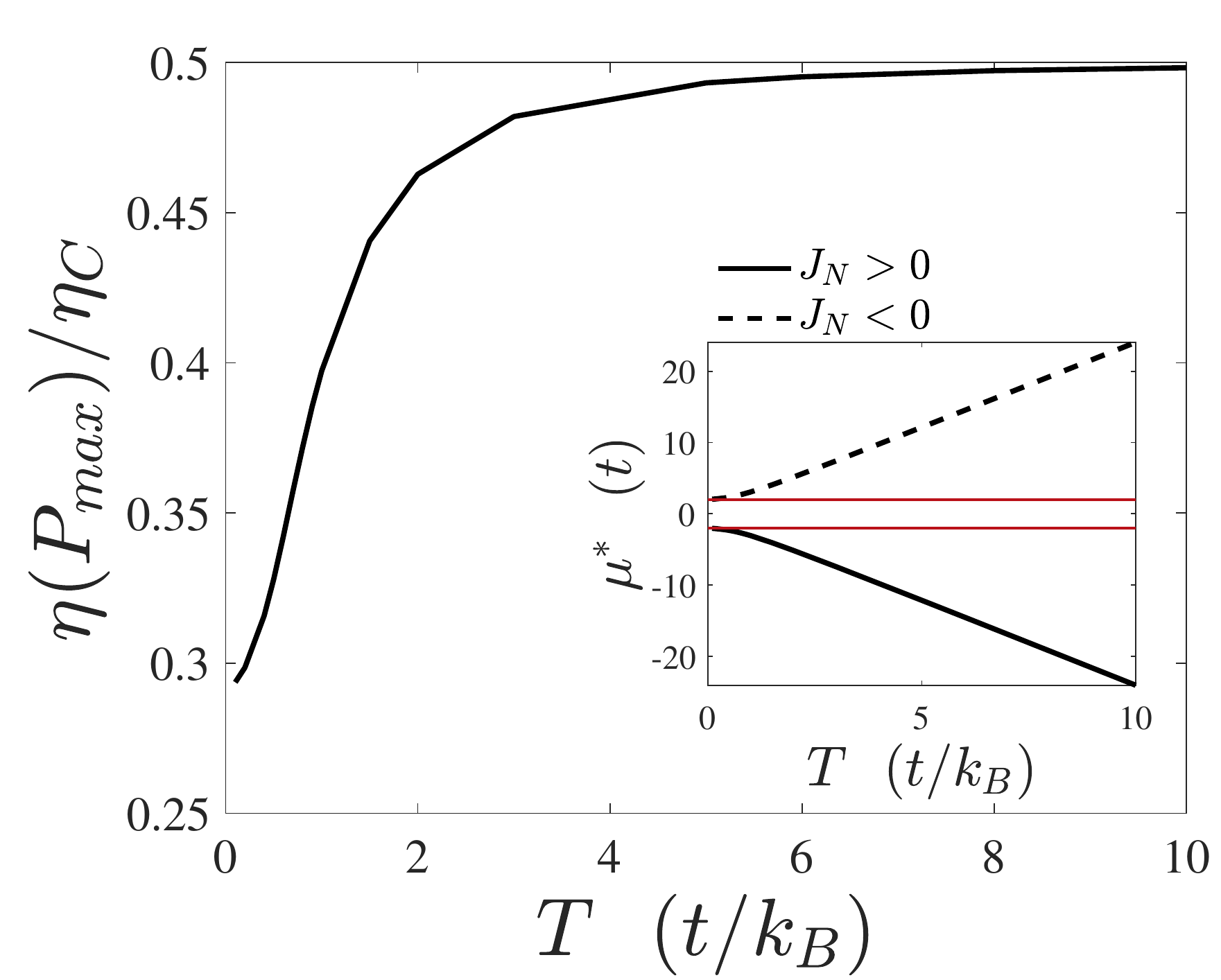} \label{fig:eff_tb}} 
\caption{(a) Absolute maximum power and (b) efficiency at maximum power, as in Fig.~\ref{fig:perf} but using a clean (i.e., nondisordered) wire as a working medium. Identical values for positive and negative current are obtained at symmetric chemical potentials relative to the center of the conducting region (inset).
}
\label{fig:perftb}
\end{figure}

\begin{figure*}
\centering
\subfloat[]{\includegraphics[scale=0.25]{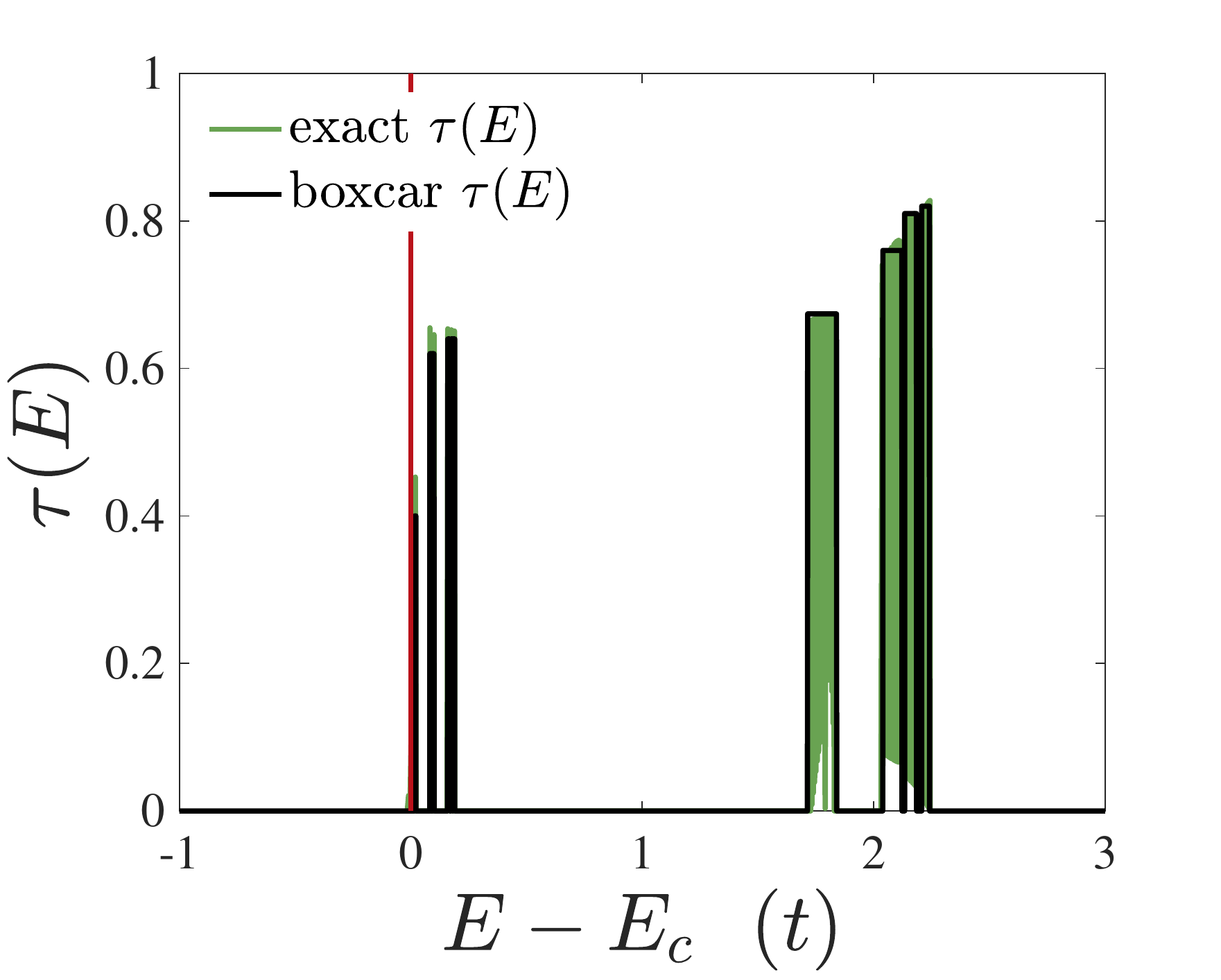} \label{fig:taubox} }
\subfloat[]{\includegraphics[scale=0.25]{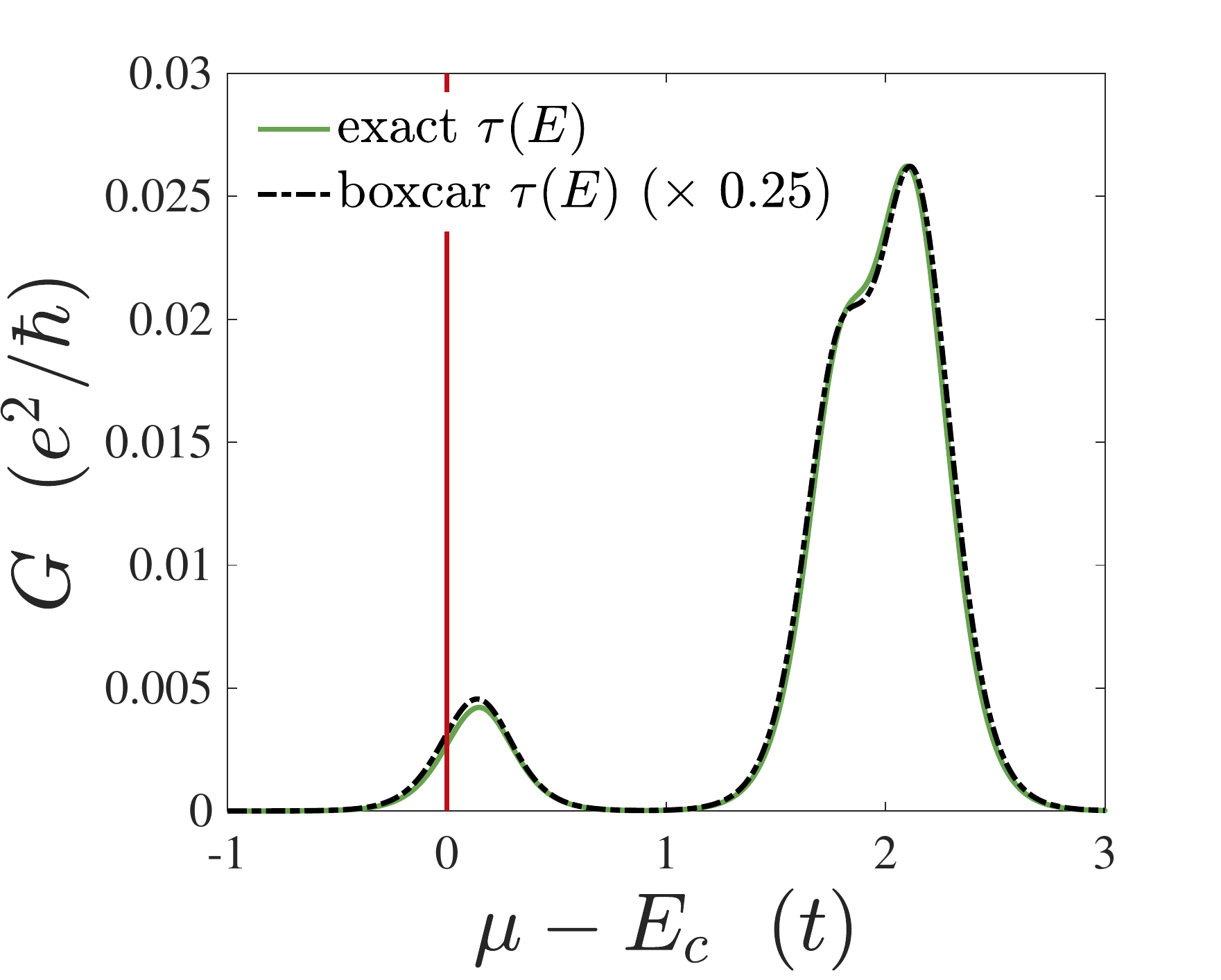} \label{fig:Gbox}}
\subfloat[]{\includegraphics[scale=0.25]{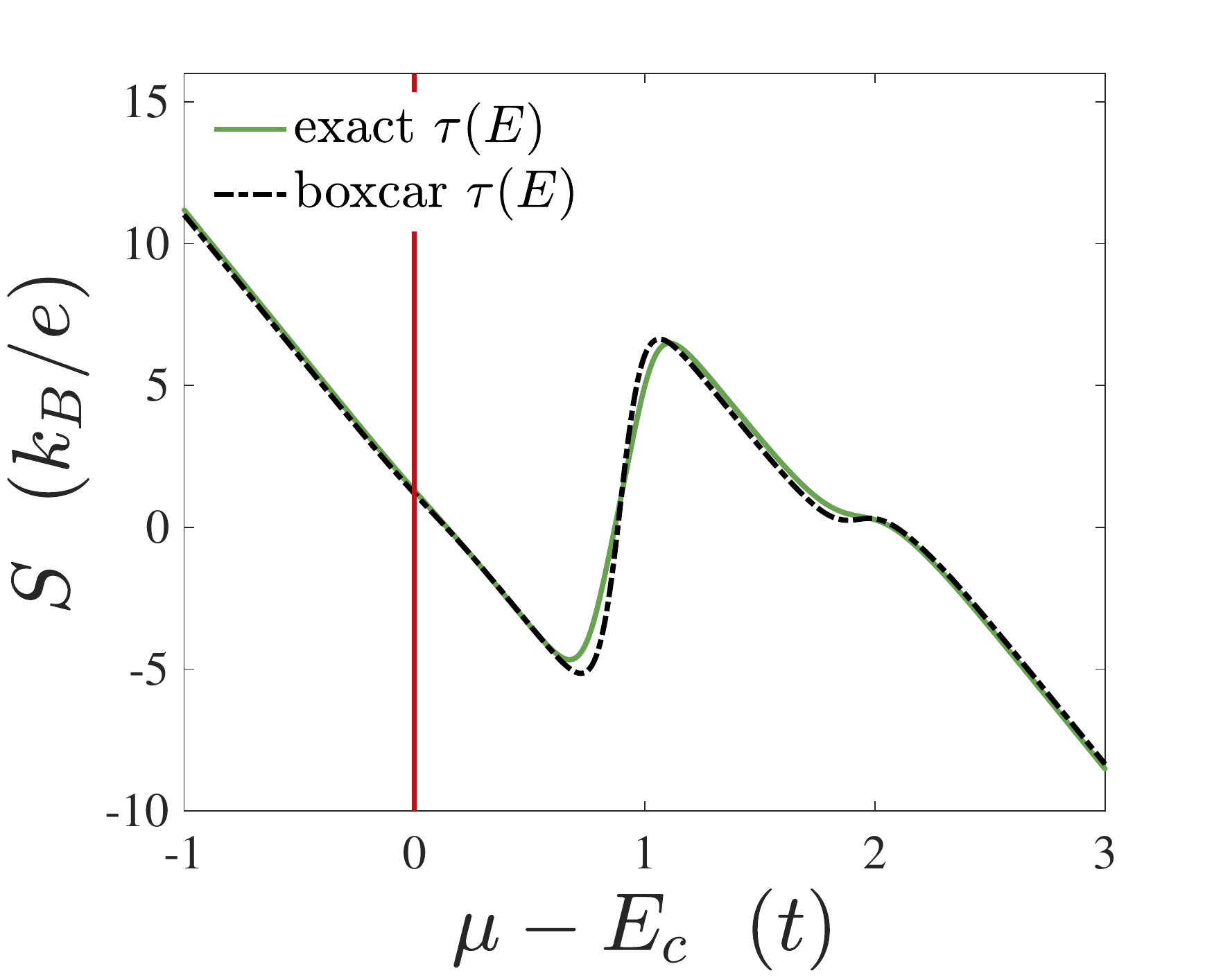}\label{fig:Sbox}}
\subfloat[]{\includegraphics[scale=0.25]{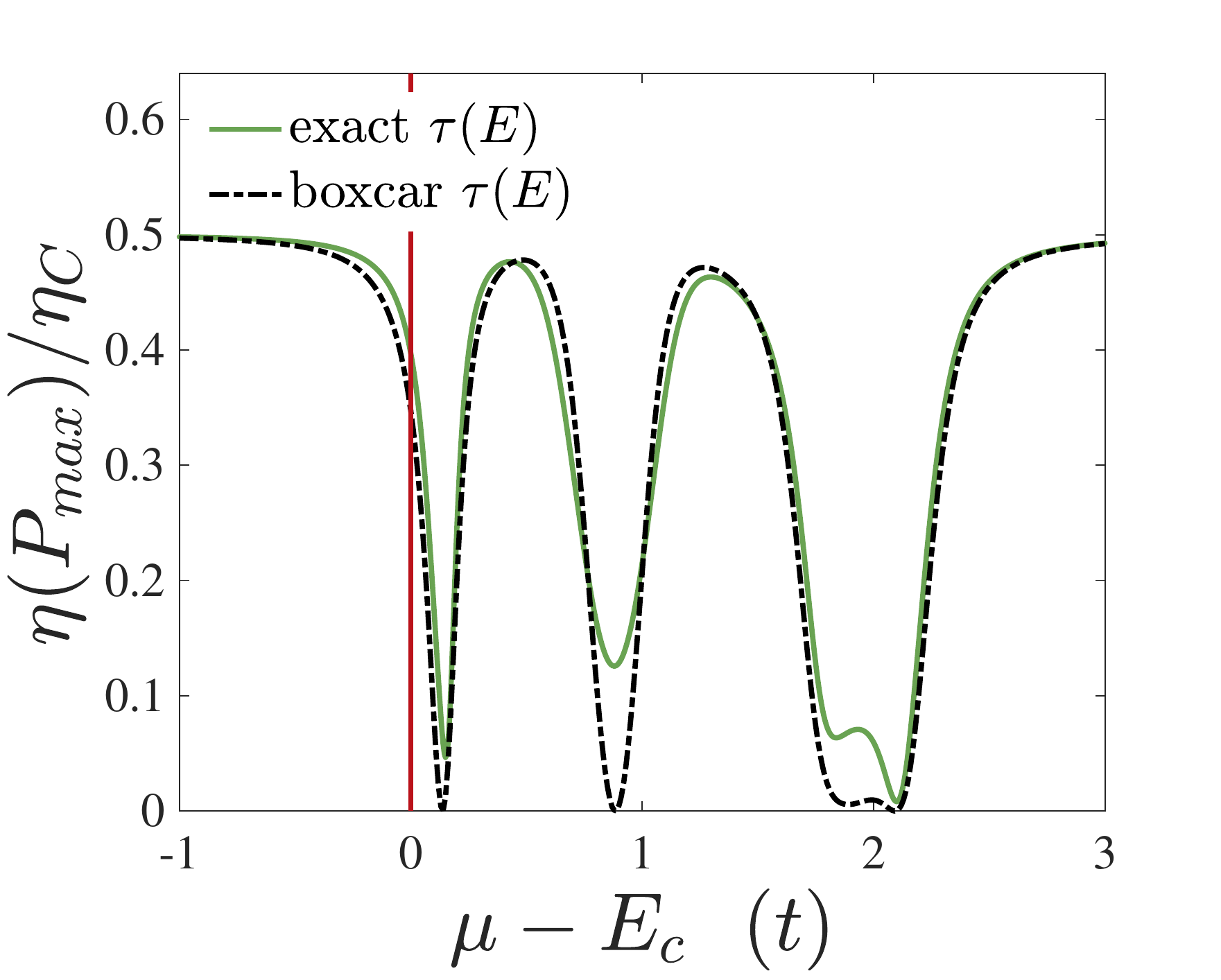} 
\label{fig:etabox}}
\caption{(a) The transmission function for the set-up in the same configuration as in the main text computed with NEGF (green lines), overlapped by a series of boxcar (black lines) following its profile. Comparison of (b) the electric conductance, (c) the Seebeck coefficient and (d) efficiency at maximum power obtained from the calculated transmission (solid green line) and from the boxcar approximation (dashed black lines).}
\end{figure*}

In Fig.~\ref{fig:perf}, we show the maximum power and the corresponding efficiency as a function of temperature. We first focus on cases where $J_N>0$. As expected, the power grows linearly with the temperature, since the thermal bias driving the currents increases as well. Furthermore, as the temperature increases, $\mu^*$ decreases, as shown in the inset of Fig.~\ref{fig:perf}. The drop in efficiency visible around $T\lesssim t/k_B$ is due to the particular structure of the spectrum: here, the transport window includes both ballistic bands, leading to a more symmetric transmission function. At even higher temperature, $\mu^*$ lies below $E_c$, where the localized states below the mobility edge are exploited to enhance the thermoelectric response (giving high efficiency), while the broad transport window still includes both ballistic bands (giving high power). Therefore, the properties of the GAAH model are here essential to obtain an efficient thermal machine with finite power output at high temperature. 

We repeat this study of maximum power for chemical potentials where $J_N<0$, shown by the dashed blue lines in Fig.~\ref{fig:perf}. The machine initially produces more power in this region, but as the temperature increases the recorded power output assumes values closer to those of the previous case and we see the two lines overlap in the plot, since the transport window broadens to covers the whole spectrum. The efficiency at maximum power converges to the CA bound, i.e., $\eta_C/2$, more quickly than in the case where $J_N>0$. Moreover, the optimal chemical potential increases with temperature, with $\mu^*$ moving well above the uppermost edge of the ballistic region for large $T$. Therefore, the strong thermoelectric response here is due mainly to the band edge.

Nevertheless, the presence of the mobility edge still enhances efficiency. In order to show this, we compute analogous data for a clean tight-binding wire, corresponding to Eq.~\eqref{wireHam} but with $V_i=0$ and  with $t=\gamma$ as before. In this case, particle-hole symmetry is broken at the edges of the spectrum located at energies $E = \pm 2t$. This leads to two perfectly symmetric points of maximum power, whose distance from the center of the spectrum at $E = 0$, one below and the other above, increases with temperature. As shown in Fig.~\ref{fig:perftb}, the efficiency saturates the CA bound at high temperature, while the maximum power is higher than for the GAAH model due to the larger number of conducting states. However, at low and intermediate temperatures where the spectral characteristics can be resolved, the thermoelectric efficiency of the clean wire, due exclusively to the presence of band edges, is lower than for a quasiperiodic system supporting a mobility edge also.

\subsection{Beyond the GAAH model: a phenomenological transmission function}

\label{sec:taubox}
The GAAH model has a fractal spectrum which is reflected by the position of the peaks in the transmission function. Here we show that the fine-grained structure of this fractal spectrum is unimportant for the physics described above. To that end, we study the transport properties of the set-up by modelling its transmission function with a series of boxcar functions of height and width corresponding to the different ballistic regions of the GAAH model. By construction, these boxcar functions lack any fine structure whatsoever.

In Fig.~\ref{fig:taubox}, the boxcar approximation is plotted together with the exact transmission function from Fig.~\ref{fig:tau1}. With this phenomenological transmission function, we now calculate the transport properties. Fig.~\ref{fig:Gbox} shows the electrical conductance $G$ as obtained from the phenomenological approach along with the exact value of $G$ for the GAAH model, showing excellent agreement up to an overall scale factor. The factor occurs because, due to the fractal nature of the spectrum of the GAAH model, the integral of the true transmission function of the GAAH model is a fraction of that of the boxcar transmission function. Other Onsager coefficients obtained from the phenomenological model also differ by the same overall factor. This, in turn, means that quantities defined as a ratio of the Onsager coefficients show excellent agreement with the GAAH model. This is shown in Fig.~\ref{fig:etabox} for the efficiency at maximum power. Due to the extreme simplicity of the phenomenological transmission function, the contribution from each boxcar function can be calculated analytically for both $G$ and $S$. Furthermore, this phenomenological transmission function can also be arrived at more microscopically in the weak system-bath coupling limit (see Appendix~\ref{app:analytical}).

This exercise shows that the physics described in previous sections is not a specific property of the GAAH model that we have considered here. Any system with similar coarse-grained features in its transmission function will show the same qualitative behaviour. Such transmission functions are expected in other quasiperiodic 1D systems with a mobility edge separating ballistic and localized states. Hence, our results exemplify the thermoelectric properties of all such systems.

\section{Conclusions}
\label{sec:conclusions}

Nanoscale thermoelectrics rely on the principle of energy filtering, where only particles within a certain energy window are allowed to flow. Here, we have shown that the localization transition in certain quasiperiodic systems gives rise to an effective energy filter and thus a novel class of efficient quantum thermal machines. In particular, we have characterized the thermoelectric properties of a generalized Aubry-Andr\`e-Harper model proposed in Ref.~\cite{GAAH1}. This model displays several remarkable spectral features, including a mobility edge in one dimension, whose position as a function of the Hamiltonian parameters is known analytically, and conducting bands lying above the mobility edge that support ballistic transport. We have shown how these properties can be exploited to design a versatile and efficient heat engine. 

These results rely on the assumption of weak applied bias, so that the response of the system is linear in the temperature and chemical-potential differences. Future work will extend the analysis to far-from-equilibrium scenarios in order to assess the full nonlinear response of the system. Here, the ability to tune the transmission profile of quasiperiodic systems by changing their Hamiltonian parameters could prove crucial in obtaining high efficiency at finite power output~\cite{whitney2014most}.

Our proposal could be experimentally tested using ultracold neutral atoms trapped in bichromatic optical lattices. Although realizing the specific GAAH potential~\eqref{eq:gaapot} may be challenging in this context, it is possible to engineer 1D quasiperiodic systems with a mobility edge by other means, e.g., by lowering the primary lattice depth so that hopping processes beyond nearest-neighbour play a role~\cite{biddle2011localization,biddle2009localization,biddle2010localization,dasSarma_bichromatic}. These systems, which have been experimentally realized recently~\cite{Bloch2019,Bloch2018}, have similar spectral features and thus should display similar thermoelectric properties to the GAAH model studied in this work. Our predictions could be tested using the toolbox of two-terminal transport measurements that has been developed for ultracold neutral atoms~\cite{Esslinger2018,Esslinger2017,Esslinger2016,Esslinger2014,Esslinger2013}. Furthermore, the effect of attractive or repulsive interparticle interactions could be experimentally investigated in such a setup. Here, a rich interplay between many-body localization, superfluidity, and nonequilibrium transport phenomena is expected to emerge.

Looking further ahead, many other families of quasiperiodic systems exist, displaying the whole gamut of possible transport behaviours (see, for example, Ref.~\cite{Fibotransport}). However, the thermoelectric properties of these systems are virtually unexplored as yet. Our work represents a first demonstration of the promise of quasiperiodic thermal machines, but their full potential for quantum thermodynamics remains to be uncovered.

\begin{acknowledgments}
This project received funding from the  European Research Council (ERC) under the European Union's Horizon 2020 research and innovation program (Grant Agreement No. 758403). We acknowledge the provision of computational facilities by the DJEI/DES/SFI/HEA Irish Centre for High-End Computing (ICHEC). J.G. and G.H. acknowledge support from the National Science Foundation under Grant No.~NSF PHY17-48958 (KITP program QTHERMO18). J.G. is supported by an SFI-Royal Society University Research Fellowship. G.H. acknowledges support from the Swiss National Foundation through the starting grant PRIMA PR00P2\textunderscore179748. 
\end{acknowledgments}

\bibliographystyle{apsrev4-1}
\bibliography{gaa_machine}

\appendix 

\section{Dependence on system-bath coupling}
\label{app:system_bath_coupling}

\begin{figure}
\centering
\subfloat[]{\includegraphics[scale = 0.24]{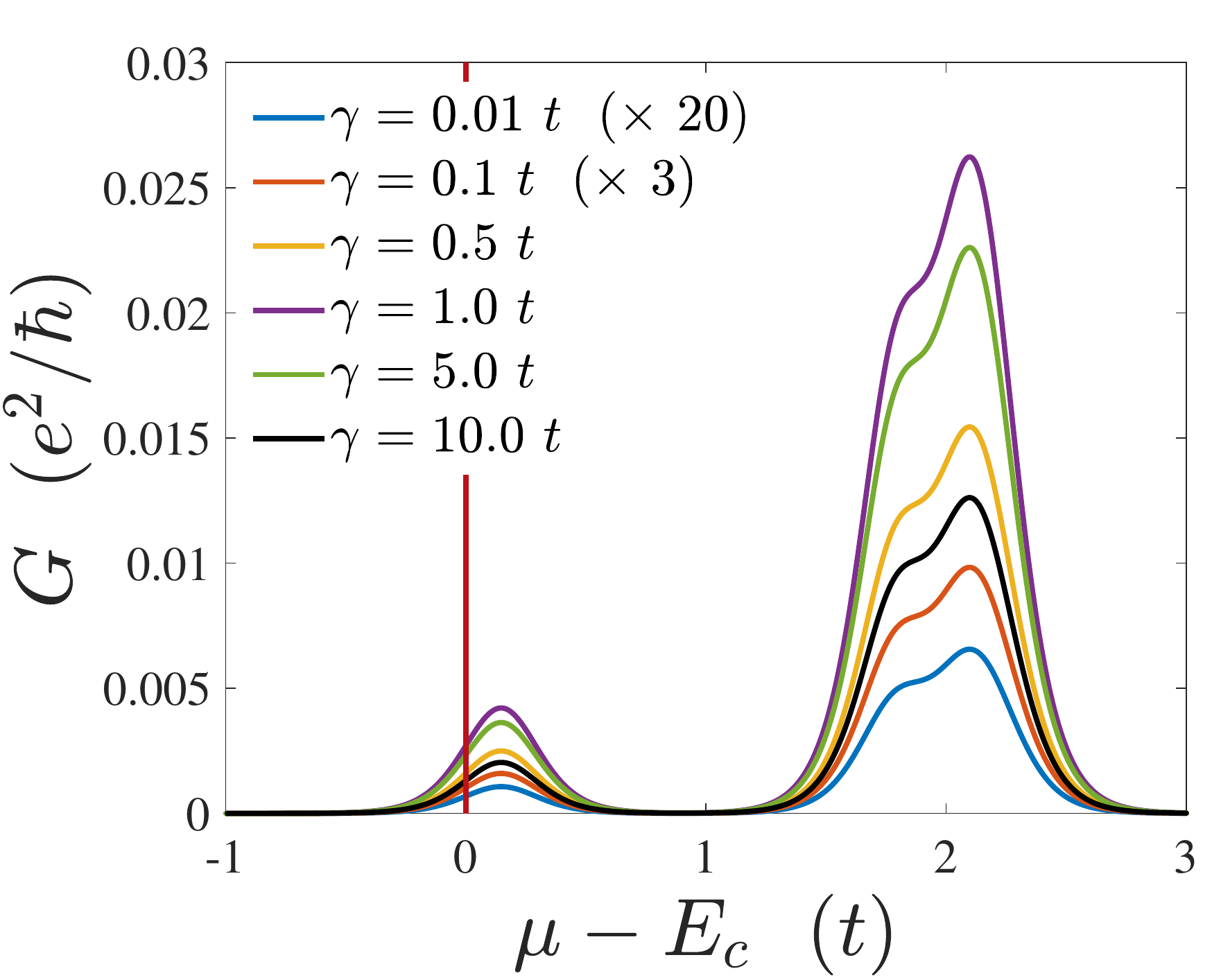} \label{fig:G_gamma}}
\subfloat[]{\includegraphics[scale = 0.24]{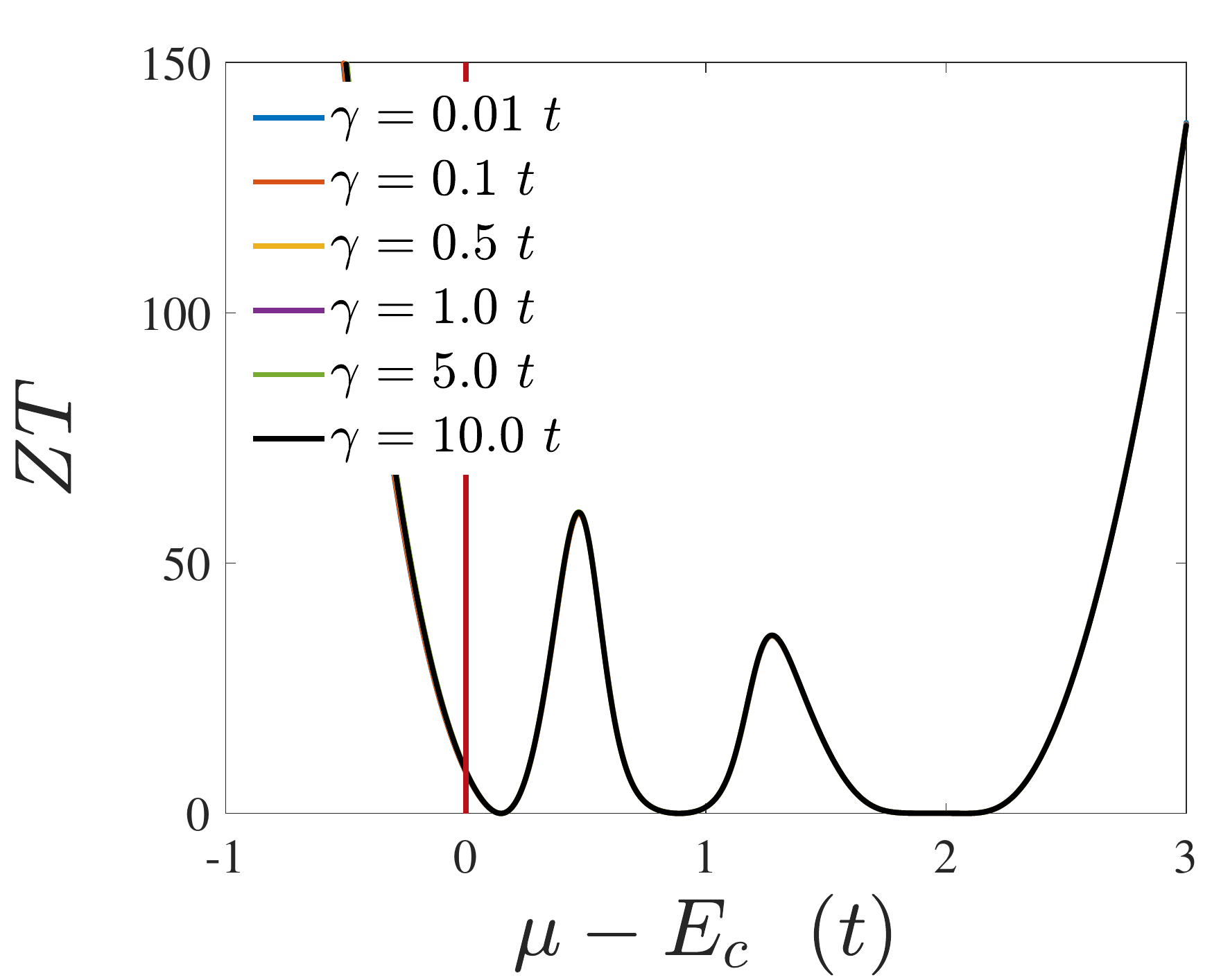} \label{fig:Z_gamma}} 
\caption{Dependence of the transport properties on the system-bath coupling strength $\gamma$. (a) Electrical conductance as a function of the distance between the chemical potential and the mobility edge, obtained for multiple choices of $\gamma$. All curves have the same form and differ only in their magnitude. (b) Figure of merit around the mobility edge for different $\gamma$. Since the three coefficients show the same behaviour, their combination is $\gamma$-independent and the different curves completely overlap. We adopt the same quasiperiodic chain used in the main text, fixing the temperature at $T = 0.1 \ t/k_B$.}
\label{fig:gamma}
\end{figure}

The study of the proposed quasiperiodic thermoelectric is characterized by a large number of parameters to control and tune in order to reach the highest possible efficiency at finite power output. In the main text we show the properties of the thermal machine computed at an intermediate coupling regime, with $\gamma$ fixed equal to $t$. The analysis of the efficiency is, in fact, independent of this choice. In order to show this, we compute the Onsager coefficients of the same system keeping the temperature and chemical potential constant and changing just the strength of the coupling between the central chain and the reservoirs. We have found that the forms of $L_{11}$, $L_{12}$ and $L_{22}$ as function of chemical potential remain the same regardless of $\gamma$, up to an overall factor. A change in the coupling constant affects just the magnitude of the coefficients, as is evident from the conductance plotted in Fig. \ref{fig:G_gamma}. The magnitude initially increases with $\gamma$, but, after reaching a maximum at an optimum $\gamma^*$, it drops as the particles begin to be scattered back to the reservoirs without entering the central region because of the high impedance mismatch. The same kind of behaviour is observed also for the other coefficients $L_{12}$ and $L_{22}$ in all temperature ranges. Quantities deriving from a ratio of the Onsager coefficients, such as the thermopower $S$ and the figure of merit $ZT$, are thus independent of $\gamma$, as shown in Fig.~\ref{fig:Z_gamma}.

\begin{figure}[b]
\centering
\subfloat[]{\includegraphics[scale = 0.24]{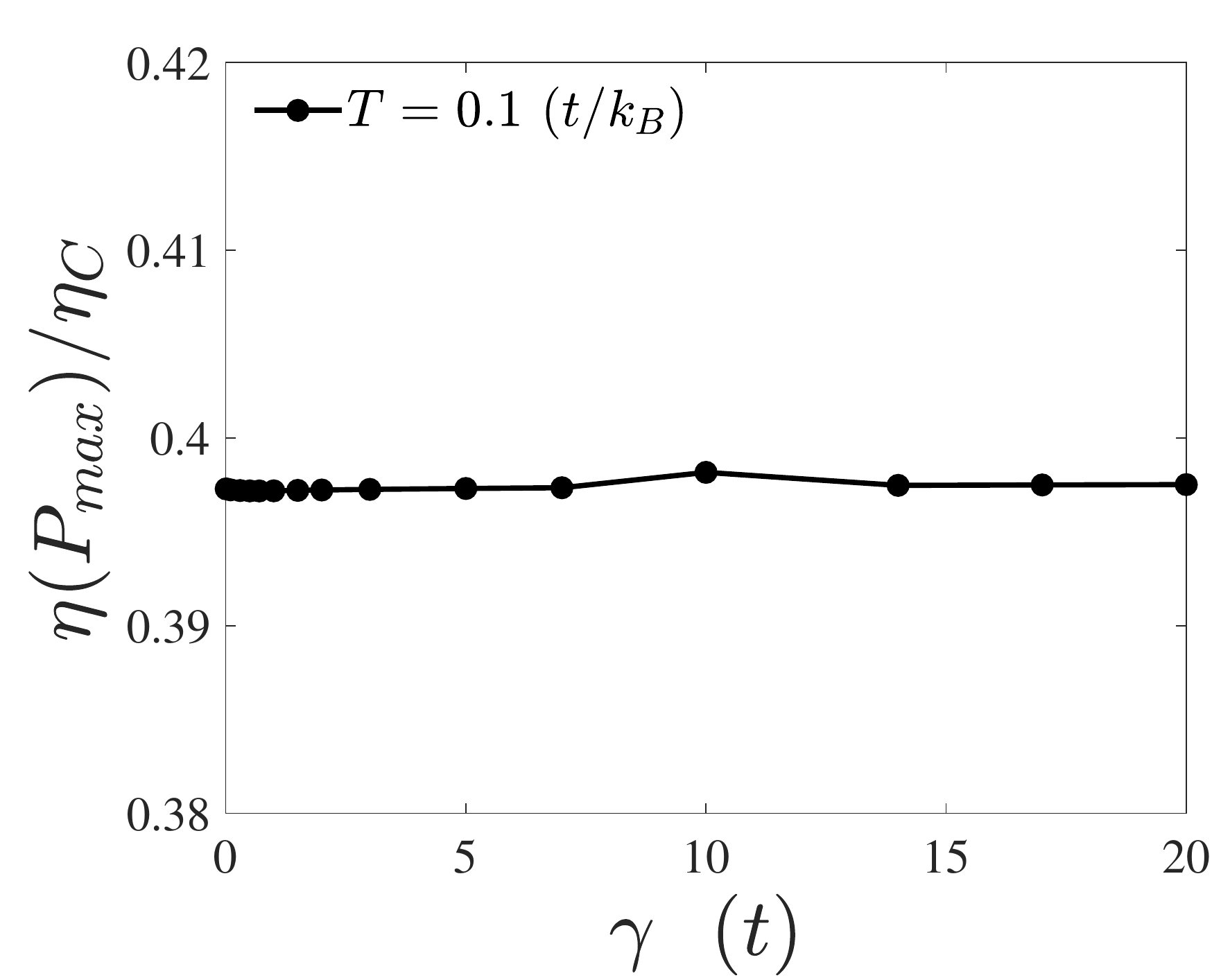} \label{fig:eta_gamma}}
\subfloat[]{\includegraphics[scale = 0.24]{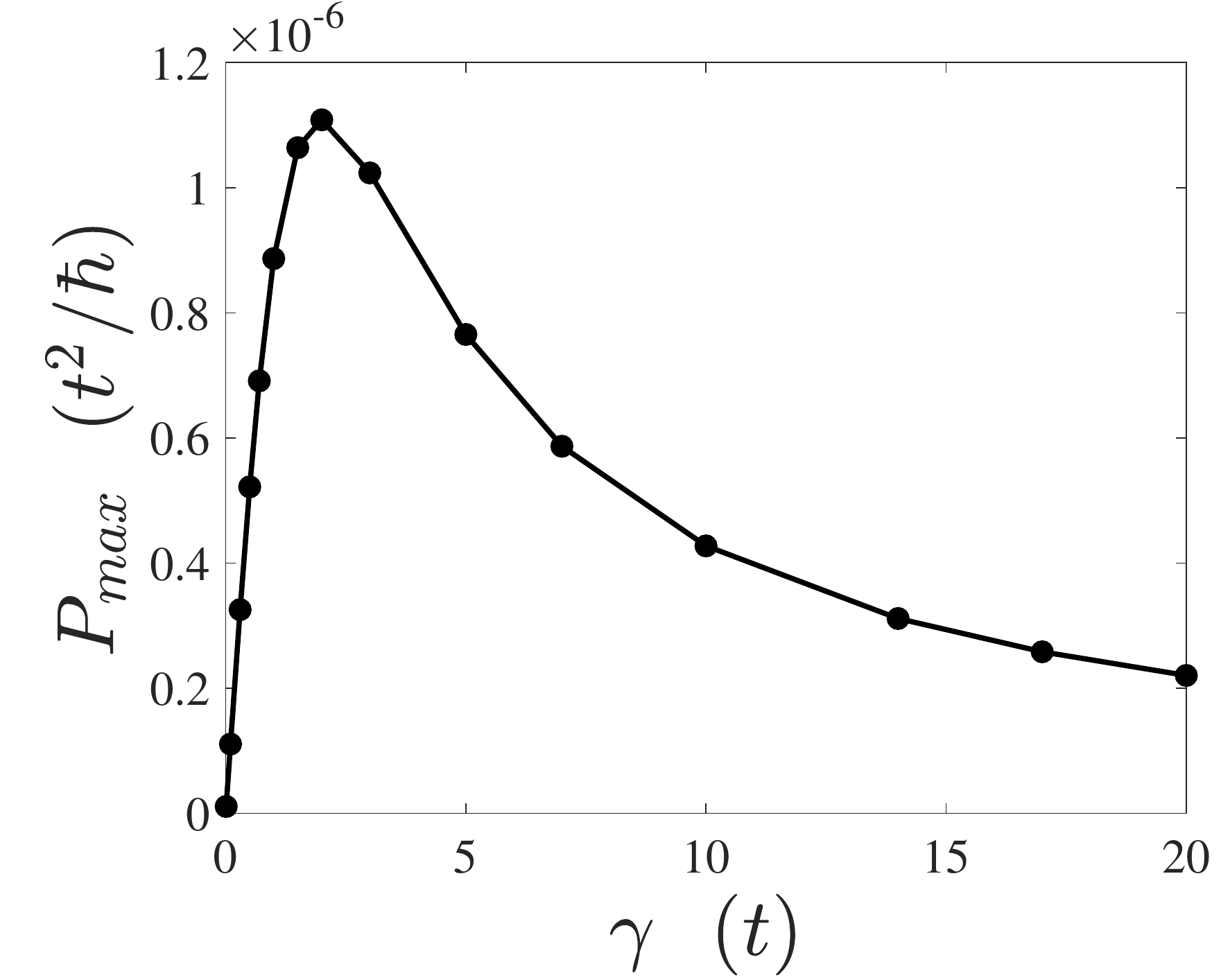} \label{fig:P_gamma}} 
\caption{(a) Efficiency at maximum power as a function of the system-bath coupling, for a fixed chemical potential, $\mu - E_c = 1.57 \ t$. Since the Onsager coefficients are modified by the same pre-factor when $\gamma$ is changed, the efficiency remains constant. (b) Maximum power transferred by the machine at the same values of $\gamma$. It is evident that it is possible to tune the system-bath coupling in such a way to optimize power without changing efficiency.}
\label{fig:gamma2}
\end{figure}

As a consequence, in the limit of large system size it is possible to maximize the power output of the machine while keeping its efficiency constant, just by tuning the coupling of the chain to the baths in the set-up.  
In Fig.~\ref{fig:gamma2} we collect, for different values of $\gamma$, results for the maximum power and the efficiency at the chemical potential which gives the highest value for the electric power output when $T = 0.1 \ t/k_B$. The corresponding value of the chemical potential is the same at every $\gamma$ and the efficiency remains constant apart from small numerical fluctuations, as expected. We see, instead, the power rising linearly for small $\gamma$, reaching the highest value at $\gamma \sim 2.0 \ t$, and subsequently decaying with a power law. The parameter $\gamma$ can be then fixed without loss of generality, and, moreover, can be used to control the maximum power output without affecting the efficiency of the thermal machine.

\section{Analytical results in the weak-coupling limit}
\label{app:analytical}

In Sec.~\ref{sec:LB} we describe the importance of breaking the symmetry between the dynamics of the electrons above and below the chemical potential, or, in other words, the electron-hole symmetry, in order to obtain a good thermoelectric. A simple way to realize this effect is to put an energy filter on the central system to prevent the transmission at certain energies. This may be achieved, for example, by a band edge or a mobility edge. Here we analytically demonstrate the enhancement of thermoelectric effects due to this mechanism. 

The particle and heat currents in a one-dimensional system among two fermionic reservoirs, within the WBL approximation and in the weak system-bath coupling  limit,  can  be  expressed  directly  as  a  function  of  the eigenstates of the isolated system~\cite{purkayastha2018}:
\begin{equation}
    J_e = 2e\gamma \sum_{n=1}^N  \frac{\Phi_{Ln}^2\Phi_{Rn}^2}{\Phi_{Ln}^2 + \Phi_{Rn}^2}  \bigl( f_L(E_n) - f_R(E_n) \Bigr),
    \label{eq:JeB}
\end{equation}
\begin{equation}
    J_q = 2\gamma \sum_{n=1}^N  \frac{\Phi_{Ln}^2\Phi_{Rn}^2}{\Phi_{Ln}^2 + \Phi_{Rn}^2}   (E_n - \mu) \bigl( f_L(E_n) - f_R(E_n) \bigr),
\end{equation}
where $\Phi_{ln}, \ l= L,R$ is the component of the $n$-th eigenstate on the first $(l=L)$ or the last $(l=R)$ site of the chain. In the linear-response regime, we thus obtain the Onsager coefficients for reference values of $\mu$ and $T$:
\begin{equation}
    L_{11} = 2\gamma e^2  T \sum_{n=1}^N  \frac{\Phi_{Ln}^2\Phi_{Rn}^2}{\Phi_{Ln}^2 + \Phi_{Rn}^2}  (-f'(E_n)),
    	\label{eq:L11B}
\end{equation}
\begin{equation}
    L_{12} = 2\gamma eT \sum_{n=1}^N \frac{\Phi_{Ln}^2\Phi_{Rn}^2}{\Phi_{Ln}^2 + \Phi_{Rn}^2}  (E_n - \mu) ( -f'(E_n) ),
    	\label{eq:L12B}
\end{equation}
\begin{equation}
    L_{22} = 2\gamma T\sum_{n=1}^N \frac{\Phi_{Ln}^2\Phi_{Rn}^2}{\Phi_{Ln}^2 + \Phi_{Rn}^2} (E_n - \mu)^2 (-f'(E_n)).
    	\label{eq:L22B}
\end{equation}
The expressions above are strictly valid only in the weak system-bath coupling regime. However, for larger $\gamma$ the only error is an overall multiplicative factor, which is the same for all the currents and Onsager coefficients. Considerations about quantities defined through ratios of Onsager coefficients can be then regarded as generic, since these prefactors cancel each other.

It is evident that in order to get a coefficient $L_{12}$ different from zero the eigenstates need to behave differently for energy above or below the chemical potential $\mu$. With this condition, the Seebeck coefficient, which is introduced in Eq.~\ref{eq:Scoeff} and enters quadratically in the definition of the figure of merit, can assume finite values.  If the spectrum of the system contains an isolated cluster of eigenstates, the strongest thermoelectric effects arises when the chemical potential is placed at their edges, since there are no states contributing below or above a certain index $n^*$ in the sum appearing in Eqs.~\ref{eq:L11B}-\ref{eq:L22B}. On the other hand, for a system exhibiting a mobility edge at $E_c = E_{n^*}$ the eigenfunctions scale with the system size $N$ as follows:
\begin{equation}
\begin{split}
\Phi_{ln}^2 &\sim e^{-N} \ \ \  \text{if} \ \ \ n < n^*, \\
\Phi_{ln}^2 &\sim \frac{1}{N}  \ \ \ \text{if} \ \ \ n > n^*.
\end{split}
\label{eq:states}
\end{equation} 
The sums can be then split into two parts: the terms for $n < n^*$ and for $n > n^*$. The former terms will go to zero as $N$ increases, while the latter will converge to a finite value.

\begin{figure}
\centering
\subfloat[]{\includegraphics[scale=0.24]{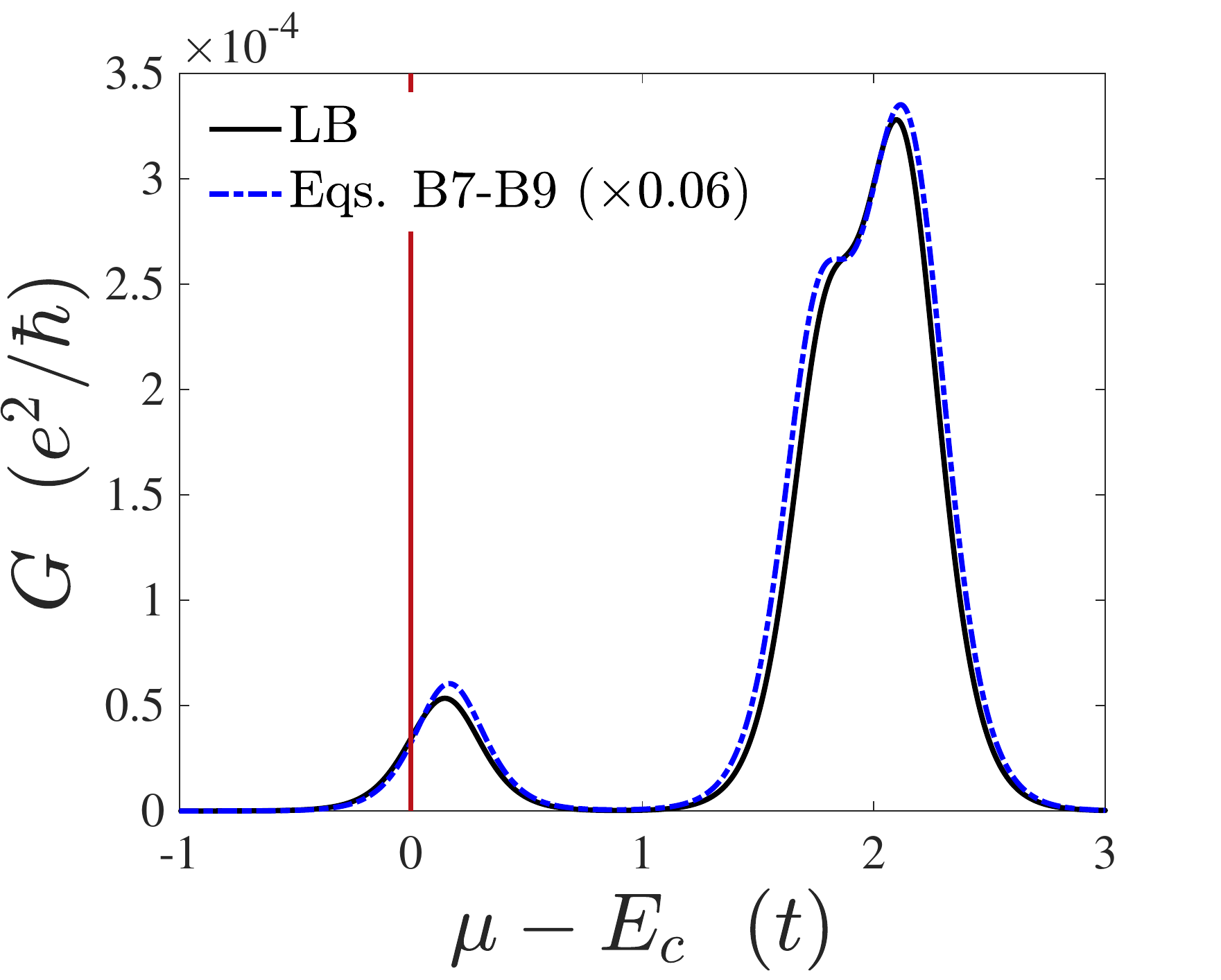} }
\subfloat[]{\includegraphics[scale=0.24]{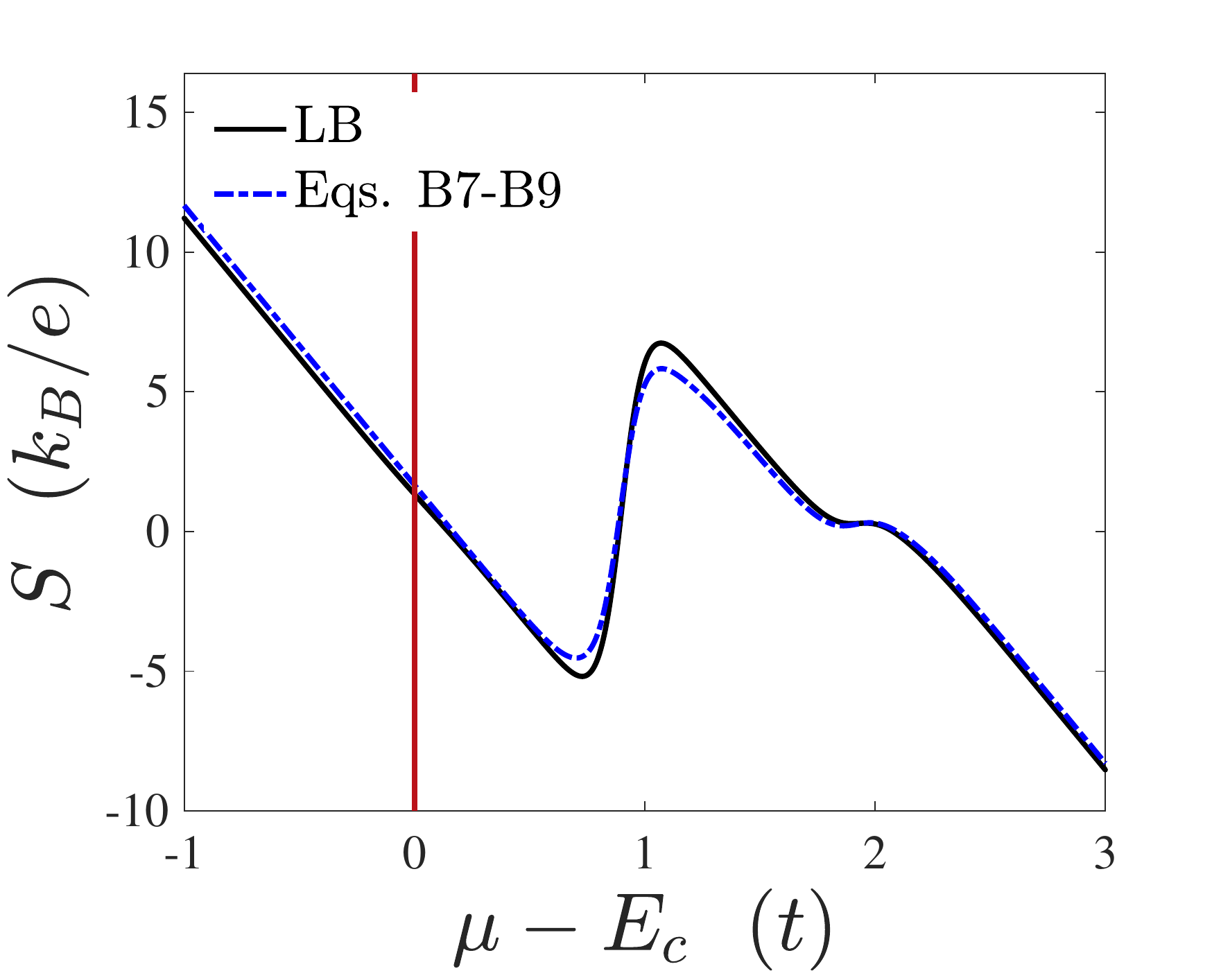} } 
\caption{ Comparison between the transport coefficients computed through the Landauer-B{\"u}ttiker integrals of Eq.~\eqref{buttik} (solid black line), and the approximated forms in Eqs.~\ref{eq:L11C}--\ref{eq:L22C} (dashed blue line). (a) Electrical conductance. (b) Seebeck coefficient. The parameters of the system are the same as in the main text, but with a weak coupling of $\gamma = 0.01 \ t$. The proportionality factor of 0.06 in (a) is a free parameter, which encapsulates the microscopic details of the eigenfunctions that are neglected in the approximations~\ref{eq:L11C}--\ref{eq:L22C}.}
\label{fig:cfr_B7}
\end{figure}

We now make a further assumption that the eigenfunctions $\Phi_{ln}$ contribute approximately the same weight for each value of $n>n^*$ in Eqs.~\ref{eq:L11B}-\ref{eq:L22B}. The Onsager coefficients for large enough $N$ can be thus approximated, up to a proportionality constant, by
\begin{align}
    L_{11} &\propto \gamma \ \frac{e^2  T}{ N}  \sum\limits_{n > n^*}^{N} (-f'(E_n)),
    \label{eq:L11C}\\
L_{12} &\propto \gamma \ \frac{e  T}{ N}  \sum\limits_{n > n^*}^{N} (E_n - \mu) (-f'(E_n)),
\label{eq:L12C}\\
L_{22} & \propto \gamma \ \frac{  T}{ N}  \sum\limits_{n > n^*}^{N} (E_n - \mu)^2 (-f'(E_n)).
\label{eq:L22C}
\end{align}
We display in Fig.~\ref{fig:cfr_B7} the comparison between the exact computation carried out through the Landauer-B{\"u}ttiker integrals and the predictions of the above equations. We notice that the proportionality constant, independent on the system size, is the same for all three Onsager coefficients. As a consequence, it does not affect quantities such as the thermopower, the figure of merit or the efficiency. Therefore, we see that we only require the single-particle eigenvalues of the system to accurately recover the essential physics, up to a proportionality constant.

\begin{figure}
\subfloat[]{\includegraphics[scale=0.24]{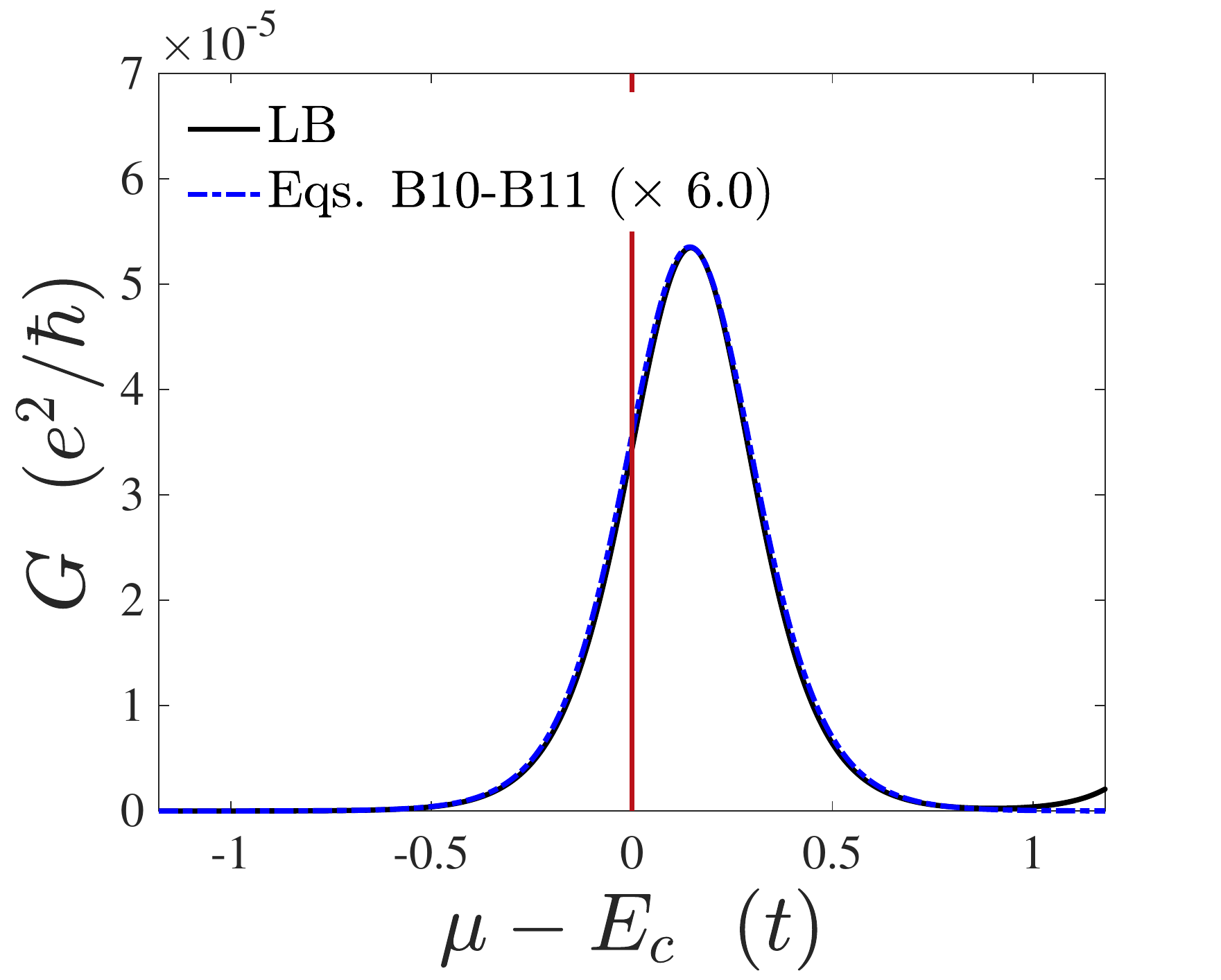} }
\subfloat[]{\includegraphics[scale=0.24]{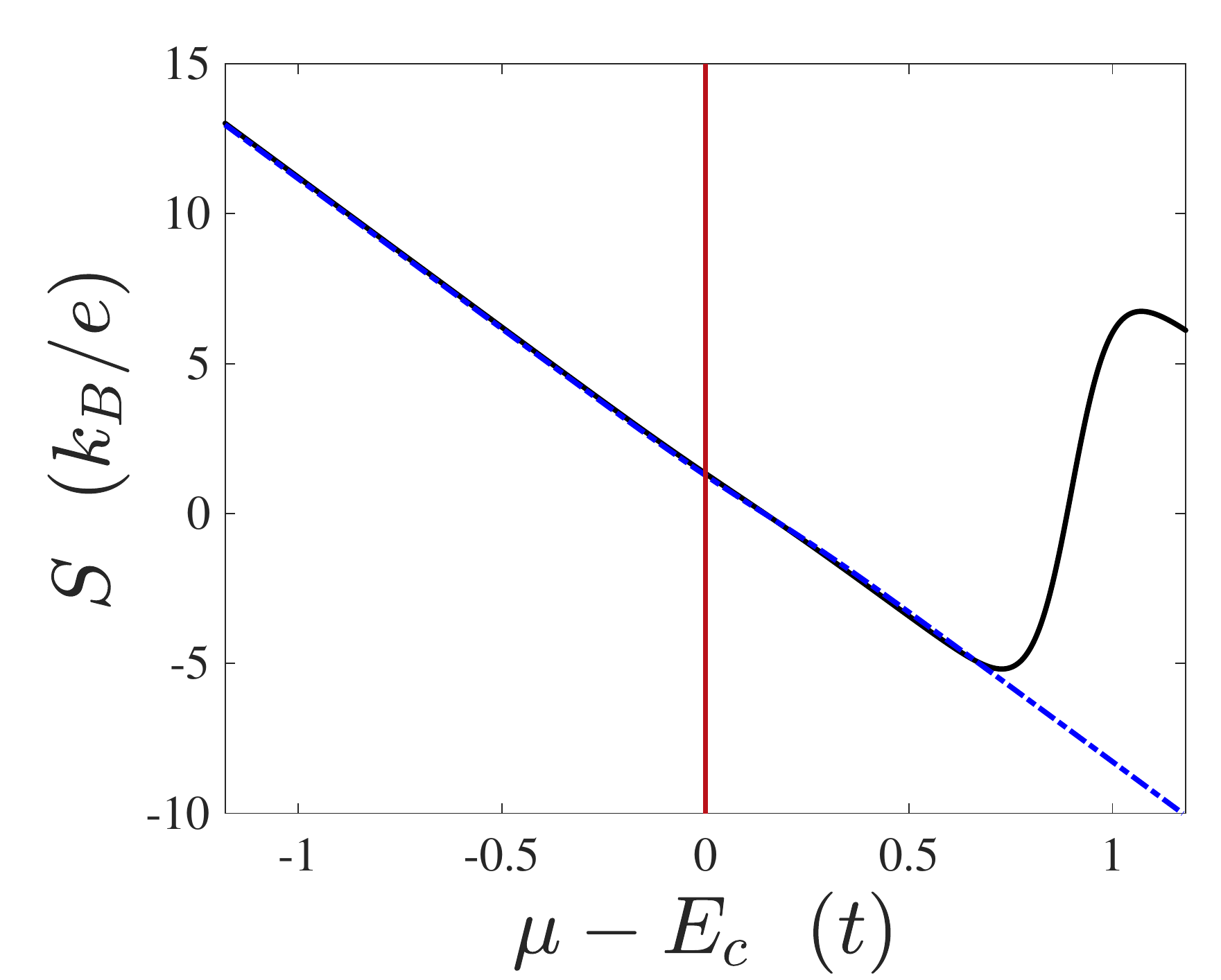} }
\caption{(a) Electrical conductance and (b) Seebeck coefficient at low temperature $T = 0.1 \ t/k_B$ and $\gamma = 0.01 \ t$. The quantities are computed through the exact Landauer-B\"uttiker integrals (solid black line), and the analytical formulae in Eqs.~\ref{eq:L11D} and \ref{eq:L12D} (dashed blue line). The proportionality factor of 6.0 in (a) is a free parameter, which reflects the fractal structure of the transmission function that is neglected in the boxcar approximation.}
\label{fig:mat}
\end{figure}

Now, we take one further step of approximation. We note that the single-particle eigenvalues occur in clusters, as evidenced by the ballistic bands in Fig.~\ref{fig:tau1}. Due to quasiperiodicity, these eigenvalue clusters have a finer self-similar structure. We now choose to completely ignore this finer structure and replace the summations in Eqs.~\ref{eq:L11C},~\ref{eq:L12C},~\ref{eq:L22C} by integrals over the width of each ballistic band. This amounts to phenomenologically modelling the transmission function by a series of boxcar functions, as done in Sec.~\ref{sec:taubox}. With this simplified assumption, we can derive closed-form analytical expressions for the contribution from each boxcar function to the Onsager coefficients $L_{11}$ and $L_{12}$. To state the result concisely, we define the following three functions:
\begin{align*}
\begin{split}
   A &= \tanh \left ( \frac{ \mu - E_1}{2 k_B T} \right), \\
   B &= \tanh\left( \frac{ E_2 - \mu}{2 k_B T} \right), \\
   C &= \log\left[ \cosh\left( \frac{\mu - E_1}{ 2 k_B T}\right) \text{sech} \left( \frac{E_2 - \mu}{ 2 k_B T}\right) \right].
\end{split}
\end{align*}
The contribution to $L_{11}$ and $L_{22}$ from a band of ballistic states between $E_1$ and $E_2$ is then given by
\begin{align}
L_{11} &\propto \frac{e^2 T \gamma}{ N} \left[ A +B \right],
\label{eq:L11D} \\
    L_{12} & \propto \frac{e T \gamma}{ N} \left[ (E_1 - \mu) A +(E_2 - \mu) B + 2 k_B T C \right]. 
    \label{eq:L12D}
\end{align}

To show the correctness of these results, we plot the conductance $G$ and Seebeck coefficient $S$ for chemical potentials $\mu$ close to the mobility edge. At low temperatures, only one cluster contributes, and this contribution should match that obtained from the above analytical formulae, up to a proportionality constant for $G$. Plots of $G$ and $S$ as obtained from the above formula are shown in Fig.~\ref{fig:mat} along with the exact results. Indeed, we see that $G$ is qualitatively identical up to a proportionality constant, while $S$ is both qualitatively and quantitatively the same. The Seebeck coefficient starts to deviate for higher $\mu$ due to contributions from the next cluster of ballistic states. This can be remedied by adding another boxcar function corresponding to the next cluster, as done in obtaining Fig.~\ref{fig:Sbox}.

\clearpage

\end{document}